\pdfoutput=1
\documentclass{aa}
\usepackage{graphicx}
\usepackage[varg]{txfonts}
\usepackage{natbib}
\usepackage[colorlinks=true, linkcolor=blue, citecolor=blue,urlcolor=blue, breaklinks=true]{hyperref}
\usepackage{dcolumn}
\begin{document}

\title{Modelling telluric line spectra in the optical and infrared with an application to VLT/X-Shooter spectra\thanks{Based on observations collected at the European Southern Observatory, Paranal, Chile, 085.C-0764(A) and 60.A-9022(C)}\fnmsep\thanks{The \texttt{tellrem} package will be available at the CDS via anonymous ftp to cdsarc.u-strasbg.fr (130.79.128.5)
or via http://cdsweb.u-strasbg.fr/cgi-bin/qcat?J/A+A/XXX/XXX as soon as the journal publishes the article.}}

\author{N. Rudolf\inst{1} \and H. M. G\"unther\inst{2,3} \and P. C. Schneider\inst{1,4} \and J. H. M. M. Schmitt\inst{1}}

\institute{Hamburger Sternwarte, University of Hamburg, Gojenbergsweg 112, 21029 Hamburg, Germany\\\email{nrudolf@hs.uni-hamburg.de} \and Harvard-Smithsonian Center for Astrophysics, 60 Garden Street, Cambridge, MA 02138, USA \and Massachusetts Institute of Technology, Kavli Institute for Astrophysics and Space Research, 77 Massachusetts Avenue, Cambridge, MA 02139 \and ESTEC/ESA, Keplerlaan 1, 2201 AZ Noordwijk, The Netherlands}

\date{Received  / Accepted }


\abstract
{Earth's atmosphere imprints a large number of telluric absorption and emission lines on astronomical spectra, especially in the near infrared, that need to be removed before analysing the affected wavelength regions.}
{These lines are typically removed by comparison to A- or B-type stars used as telluric standards that themselves have strong hydrogen lines, which complicates the removal of telluric lines. We have developed a method to circumvent that problem.}
{For our IDL software package \texttt{tellrem} we used a recent approach to model telluric absorption features with the line-by-line radiative transfer model (LBLRTM). The broad wavelength coverage of the X-Shooter at VLT allows us to expand their technique by determining the abundances of the most important telluric molecules H$_{2}$O, O$_{2}$, CO$_{2}$, and CH$_{4}$ from sufficiently isolated line groups. For individual observations we construct a telluric absorption model for most of the spectral range that is used to remove the telluric absorption from the object spectrum.}
{We remove telluric absorption from both continuum regions and emission lines without systematic residuals for most of the processable spectral range; however, our method increases the statistical errors. The errors of the corrected spectrum typically increase by 10\,\% for $S/N\sim10$ and by a factor of two for high-quality data ($S/N\sim100$), i.e. the method is accurate on the percent level.}
{Modelling telluric absorption can be an alternative to the observation of standard stars for removing telluric contamination.}


\maketitle

\begin{table*}
\caption{List of objects and observational details for the objects discussed in this paper.}
\label{tab:obsdetails}
\centering 
\begin{tabular}{lrlrr|rrr|rrr}
\hline\hline
Object & $m_{J}$\tablefootmark{a} & \multicolumn{1}{c}{Start of observation} & \multicolumn{1}{c}{Mean} & \multicolumn{1}{c|}{Mean} & \multicolumn{3}{c|}{Exposure time [s]\tablefootmark{b}} & \multicolumn{3}{c}{Pipeline provided $S/N$} \\
 & & \multicolumn{1}{c}{(UTC)} & seeing [$^{\prime\prime}$] & airmass & UVB & VIS & NIR & UVB\tablefootmark{c} & VIS\tablefootmark{d} & NIR\tablefootmark{e}\\
\hline
\object{EG 274} & 11.6 & 2010-May-05 03:24:44 & 0.68 & 1.26 & 1$\times$90 & 1$\times$120 & 1$\times$120 & 299 & 190 & 71\\
\object{MV Lup} & 9.6 & 2010-May-05 08:30:13 & 0.88 & 1.32 & 2$\times$150 & 4$\times$60 & 6$\times$100 & 268 & 293 & 377\\
\object{V895 Sco} & 10.1 & 2010-May-05 07:50:21 & 0.98 & 1.12 & 2$\times$300 & 4$\times$140 & 6$\times$150 & 140 & 241 & 390\\
\object{S CrA} &  8.2 & 2010-May-05 09:48:34 & 1.02 & 1.07 & 2$\times$150 & 4$\times$30 & 6$\times$50 & 363 & 294 & 791\\
\hline
\end{tabular}
\tablefoot{\tablefoottext{a}{$m_{J}$ from 2MASS \citep{2MASS}}; \tablefoottext{b}{except for EG~274 all observations were carried out in nodding mode, N$\times$ states the number of exposures taken in total at the different nodding positions}; \tablefoottext{c}{in region 4500\,--\,4700\,\AA{};} \tablefoottext{d}{in region 7450\,--\,7560\,\AA{};} \tablefoottext{c}{in region 15\,140\,--\,15\,480\,\AA{}.}}
\end{table*}

\section{Introduction}\label{sec:intro}
The strong and variable absorption of Earth's atmosphere interferes with almost
all ground-based astronomical observations. 
In the optical and infrared spectral range the flux of a celestial body
is  diminished by extinction; in addition,  so-called telluric absorption lines are 
imprinted on the spectra, a fact particularly relevant for
high-resolution spectroscopy.
The number of telluric lines depends on wavelength and  
for the specific case of the VLT/X-Shooter with its
wavelength coverage ranging from 3100\,\AA{} to 24\,700\,\AA{}, 
starting from around 6000\,\AA{} bands of water (H$_{2}$O) and molecular oxygen (O$_{2}$) are predominantly present, accompanied by large 
contributions of carbon dioxide (CO$_{2}$) above 12\,000\,\AA{} and of 
methane (CH$_{4}$) above 16\,000\,\AA{}. As is well known, the atmospheric transmission already becomes quite small in some restricted wavelength bands  in the
infrared range.

The typical procedure used to remove telluric absorption features and recover the original spectrum is to observe a standard star close in time and airmass to the object of interest. 
The spectrum of the object is then divided by the spectrum of the standard star. Early-type stars of spectral type B or A are usually chosen as standard stars because their spectra show   few and rather weak metal lines. 
Unfortunately, these stars have strong intrinsic photospheric hydrogen absorption features that cause severe difficulties if  the Paschen and Brackett lines of the science object are the points  of  interest.
The simplest solution for this problem would be to interpolate over the hydrogen absorption and accept the resulting uncertainties.

Several more sophisticated methods have been developed: (a) using solar-like stars as telluric standards and a high-resolution solar spectrum \citep{1996AJ....111..537M}; (b) combining   this first method with the standard early-type star method and the fitting of the hydrogen line profiles \citep{1996ApJS..107..281H,2005ApJS..161..154H}; and (c) using A0 stars as telluric standards and a high-resolution model spectrum of Vega \citep{2003PASP..115..389V}.  Despite its wide adoption in the literature, using any astrophysical source as a reference  telluric standard has some fundamental limitations (see discussion in \citet{2010A&A...524A..11S}): First, it takes away valuable observing time from the science targets of an observation and second, its accuracy is limited by how well we know the intrinsic spectrum of the calibration star. Often it is sufficient to perform a good calibration in certain regions of the spectrum, and A and B stars can work very well for this purpose unless the flux and shape of hydrogen lines is critical to the scientific analysis. Despite all the advances in spectral modelling, there are uncertainties in the reference spectrum, not  least because individual stars might differ in rotational velocity, age, and metallicity. However, telluric standard stars need to be observed at the same airmass and close in time to the science target, so they need to be chosen from a small region on the sky close to the science target. Unfortunately, this might force the observer to select a less well characterised star as the telluric standard. In principle, the same limitations apply to all the variations of this technique mentioned above. For example, when using a solar-type  telluric standard, the regions around hydrogen lines might be well calibrated since these lines are weak in solar-like stars, but any spectral region where the Sun has photospheric features might be inaccurate if the star chosen as  calibrator does not match the solar abundance, period, or mass exactly.

An alternative approach to the observations of telluric standards is the use of theoretical models of the atmospheric transmission. This idea has been successfully pursued, for example by \citet{1993A&A...271..734L}, \citet{1994A&A...282..879W}, \citet{2007PASP..119..228B}, or \citet{2010A&A...524A..11S}. The technique to remove telluric absorption presented by \citet{2010A&A...524A..11S} was used by e.g. \citet{2010ApJ...713..410B} and \citet{2012A&A...538A.141R}. Based on their technique we developed a method for removing telluric lines from VLT/X-Shooter spectra. In this paper we explain the details of this method and evaluate its accuracy and its contribution to the overall error budget. This paper is accompanied by a software package that implements the method described here.

Our paper is structured as follows. In Sect. \ref{sec:obsdatared} we describe the observations taken with VLT/X-Shooter and their reduction. In Sect. \ref{sec:specmod} we describe the spectral modelling of the telluric absorption. We restrict ourselves to describing the general functionality of the package and refer to the user manual accompanying the package for details on its use, which we demonstrate in Sect. \ref{sec:results}. We present a summary and our conclusions in Sect. \ref{sec:sumcon}.

\section{Observations and data reduction}\label{sec:obsdatared}
Our specific aim is the application of the method to spectra of classical T Tauri stars (CTTS) observed with the VLT/X-Shooter instrument \citep{2011A&A...536A.105V} and  in this paper we present example spectra to highlight the benefits of our new method. Details of the observations can be found in Table~\ref{tab:obsdetails}.

Our observations were carried out in visitor mode during the nights of 2010 May 4 and 5. X-Shooter is a multiwavelength medium-resolution spectrograph consisting of 3 arms; each arm is an independent cross-dispersed echelle spectrograph equipped with optimised optics, dispersive elements, and detectors. The UVB arm covers the wavelength range between
3100\,--\,5900\,\AA{}, the VIS arm  between 5400\,--\,10\,100\,\AA{}, and the NIR arm between 9900\,--\,24\,700\,\AA{}. The objects were observed in nodding mode with the 0.5$^{\prime\prime}\times 11^{\prime\prime}$ slit in the ultraviolet (UVB) arm and the 0.4$^{\prime\prime}\times 11^{\prime\prime}$ slit in the visual (VIS) and the near-infrared (NIR) arms. This instrumental set-up results in a resolution of $R\sim10\,000$ in the UVB arm, $R\sim18\,000$ in the VIS arm, and $R\sim10\,000$ in the NIR arm. Additionally, the flux standard star EG~274, which is a white dwarf, was observed in offset mode with the 5.0$^{\prime\prime}\times 11^{\prime\prime}$ slit in all arms. For all observations the slit was set to the parallactic angle. Object acquisition was done in the optical using a $V$ filter, telescope tracking was done at 4700\,\AA{}.

The data reduction was carried out with the X-Shooter pipeline \citep{2010SPIE.7737E..56M} version 1.5.0 for the nodding mode for each spectrographic arm individually, following the standard steps that include bias subtraction, sky subtraction via differencing the images from the nodding position A and B, flat-fielding, order-tracing and merging, wavelength calibration, and extraction. In offset mode a pair of images is taken, one with the object and one with sky only, and sky subtraction is accomplished by differencing. However, instead of using the pipeline flux calibration we carried out our own procedures. For the flux standard star EG~274 we first removed the telluric lines from the spectrum using the method described in this paper. We then determined the response curve by dividing the corresponding spectrum from the X-Shooter spectrophotometric standard stars catalogue \citep{1994PASP..106..566H,2010HiA....15..535V} by the observed telluric line corrected spectrum normalised to 1\,s exposure time. This ratio was smoothed to reduce the noise using a sliding box-car algorithm with a box  width of 50 spectral bins ($\approx 20$~\AA) in the UVB and VIS arms and 200 bins ($\approx200$~\AA) in the NIR arm and used for flux calibration.

\section{Spectral modelling of Earth's atmosphere}
\label{sec:specmod}

In order to compute the theoretical transmission of Earth's atmosphere we apply the technique presented by \citet{2010A&A...524A..11S}. Here we provide an overview of this technique and refer to their paper for a more detailed description.  Basically, the transmission of Earth's atmosphere is computed
using the radiative transfer code line-by-line radiative transfer model (LBLRTM)\footnote{http://rtweb.aer.com/lblrtm\_frame.html}, which in turn
is based on  the fast atmospheric signature code \citep[FASCODE;][]{1992JGR....9715761C,2005JQSRT..91..233C}. The LBLRTM requires a model of Earth's atmosphere and line data as input; all the atmospheric data is available independent of the astronomical observation and can be retrieved from public meteorological databases.

The required model of the atmosphere provides the vertical temperature, pressure, and molecular abundance profiles. 
It is built by combining two models. The meteorological models from the Air Resources Laboratory (ARL) at the US National Oceanic and Atmospheric Administration (NOAA) provide temperature, pressure, and dew point temperature for surface heights $\leq$ 26\,km. The dew point temperature serves as the density measure for H$_{2}$O in the atmosphere. These so-called sounding files are available at the Global Data Assimilation System (GDAS)\footnote{http://ready.arl.noaa.gov/READYamet.php} in 3\,h intervals. For each object we use the model closest in time to the observation. Temperature, pressure, and density for surface heights above 26\,km for H$_{2}$O and for all surface heights for the other molecules are provided by an equatorial model atmosphere\footnote{http://www-atm.physics.ox.ac.uk/RFM/atm/} from the Michelson Interferometer for Passive Atmospheric Sounding (MIPAS), constructed by John Remedios (U.\ Leicester).
The required line data comes from the high-resolution transmission molecular absorption database (HITRAN) in its 2008 edition \citep{2009JQSRT.110..533R}, which provides frequency, line strength, and pressure broadening coefficients for spectral lines of 42 different molecules.

To compute a telluric model spectrum for a specific observation taken at an observation altitude $h$ with the required accuracy, the total molecular abundances of H$_{2}$O, O$_{2}$, CO$_{2}$, and CH$_{4}$ in the atmospherical model need to be adjusted by fitting. The full model spectrum thus consists of three components: (1) the stellar spectrum, which is multiplied with (2) the atmospheric transmission spectrum and then convolved with (3) the instrumental profile.

Of course, the stellar spectrum is not known a priori. One could thus devise an iterative procedure where we start with a simple guess for the stellar spectrum, fit the telluric abundances and the instrumental profile, and then use this information to extract the stellar spectrum which would serve as the starting point for the next iteration. Fortunately, the signatures of the telluric molecules are very distinct and a very simple description for the stellar spectrum is sufficient to fit the telluric lines.

In principle, all these parameters could be fit for the entire spectrum at once, but in practice we found it much easier to perform the fit in only one small wavelength interval at a time because this is much faster to compute and LBLRTM can only compute a wave number segment of 2020\,cm$^{-1}$ in one run ($\sim$ 1000\,\AA{} around 7000\,\AA{}, $\sim$ 5300\,\AA{} around 16\,000\,\AA{}). We use wavelength intervals of 300\,\AA{}. Over this short range, the stellar spectrum can be approximated as a straight line for all types of objects in our sample (CTTS, white dwarf, B stars). In fact, the method works well even in the presence of absorption and emission lines unless they dominate the fit instead of the telluric lines. The instrumental profile of X-Shooter is represented by a Gaussian where we fit the width. We also allow for a shift in wavelength, which is necessary to correct for small inaccuracies in the wavelength calibration. The measured shifts could in principle be used to test and improve the wavelength calibration. Both instrumental parameters vary over the wavelength range of X-Shooter. The wavelength shift changes because of changes in the accuracy of the wavelength calibration along the different X-Shooter orders, which lead to slightly different wavelength shifts of a few tenths of \AA;{}  the widths of the Gaussian needed to adjust the telluric transmission spectrum to the observed spectrum change because of the different resolutions in the arms. Within segments of 300\,\AA{} length we found the wavelength calibration and the width of the profile to be stable.

Each molecule in the Earth's atmosphere has very distinct spectral bands where the absorption lines from this single molecule dominate the total telluric absorption spectrum. We make use of the broad wavelength coverage of X-Shooter and the fact that the abundances do not depend on wavelength. We searched for regions in the spectral range of X-Shooter that are almost free of stellar emission features and dominated by telluric lines, but might still contain some stellar absorption features. We found several spectral regions where only one of the main contributors to the telluric signal generates lines or the only other contributor is H$_{2}$O. They are listed in Table~\ref{tab:molabdet}.

\begin{table}
\caption{Spectral ranges used to determine the abundances of H$_{2}$O, O$_{2}$, CO$_{2}$, and CH$_{4}$.}
\label{tab:molabdet}
\centering 
\begin{tabular}{ll}
\hline\hline
Wavelength range [\AA{}] & Molecule fitted\\
\hline
7590--7660 & O$_{2}$\\
9040--9210 & H$_{2}$O\\
9270--9510 & H$_{2}$O\\
12\,550--12\,775 & O$_{2}$\\
13\,160--13\,280 & H$_{2}$O\\
14\,470--14\,685 & H$_{2}$O\\
14\,685--14\,900 & H$_{2}$O\\
19\,950--20\,300 & CO$_{2}$\\
20\,450--20\,800 & CO$_{2}$\\
22\,850--23\,110 & CH$_{4}$\\
23\,370--23\,630 & CH$_{4}$\\
23\,630--23\,900 & CH$_{4}$\\
\hline
\end{tabular}
\end{table}

First, we investigate the regions where H$_2$O is the only significant contributor. For each of these regions we fit the water abundance in the Earth's atmosphere, the Gaussian width of the instrumental profile, the wavelength shift, and the slope and intercept of the straight line that represents the stellar continuum. We keep the remaining abundances fixed at some arbitrary value because the other molecules do not contribute to the telluric lines in this region.
The abundances found for each region typically differ by less than 10\,\%. We use the mean of these values as the abundance of water for all following computations.

We found it advantageous to determine the abundances of the various contaminants successively. We thus fit the water abundance first, since water lines appear at many wavelengths. With the given water abundance, we identify wavelength regions that contain O$_{2}$ and fit for the O$_{2}$ abundance keeping the water abundance fixed. The remaining molecules (i.e. CO$_{2}$ and CH$_{4}$) are treated in a similar manner. For each selected wavelength region (see Table~\ref{tab:molabdet}) we fit the abundance of the respective molecule and the width of the instrumental profile, a wavelength shift, and the slope and intercept of the line that represents the stellar continuum. For these fits we use the water abundance determined above and keep the remaining molecules that do not contribute to the telluric absorption in this region fixed. Finally, we compute the mean of the measured abundances for each molecule, which typically agree to within 10\,\%.

With the molecular abundances known, we finally step through the entire observed spectrum in 300\,\AA{} intervals, fitting the width of the instrumental line profile, the wavelength shift, and the slope and intercept of the straight line that represents the continuum. There is no need to obtain an accurate model for the stellar spectrum as we are only interested in the instrumental profile and the wavelength shift at this point, both of which are seen in the telluric lines. We find that our very simple approximation for the stellar spectrum is sufficient to avoid fitting artefacts. Describing the stellar spectrum with a straight line and not merely by a constant takes into account that the spectrum already may show a significant change over 300\,\AA{}. Thus we avoid fitting the width of the instrumental line profile  with a very large value in cases where there are a few telluric lines on one side of the spectrum, but not on the other where the fit attempts to compensate for the seemingly missing flux if only a constant is used to approximate the stellar spectrum. In principle,  this problem  can also be circumvented by normalising the observed spectrum prior to the fit. This approach would also work for line dominated spectra, i.e. when the spectrum cannot be reasonably represented by a straight line even on short wavelength ranges. Here, however, we prefer a forward-modelling approach where the data remains unchanged.

For each segment we correct for the telluric absorption by dividing the observed spectrum by the modelled telluric spectrum, convolved with the instrumental line profile, and shifted in wavelength according to the fit. We scale the uncertainties on the observed spectrum that were derived by the data reduction pipeline in the same way.

For all fits we use the least-squares fitting code MPFIT \citep{2009ASPC..411..251M} to minimise the difference between the observed and the telluric model spectrum. 
On a standard desktop PC the computation time needed  to remove the telluric lines from one X-Shooter spectrum is about 15\,min.

\section{Results}\label{sec:results}
We applied the method described in Sect.~\ref{sec:specmod} to a spectrum of the white dwarf EG 274 and several spectra of CTTS obtained with X-Shooter. Telluric absorption lines are only visible in the VIS and NIR arms of X-Shooter, telluric emission lines are removed by the nodded observation mode. The regions between 13\,400\,\AA{} and 14\,500\,\AA{} and between 18\,100\,\AA{} and 19\,600\,\AA{} cannot be corrected with this method because they are too heavily contaminated to obtain a reasonable fit of the telluric lines (see Fig.~\ref{fig:completespecwm}). This divides the spectrum of the NIR arm  into three parts ranging from 9900\,--\,13\,400\,\AA{}, 14\,500\,--\,18\,100\,\AA{}, and 19\,600\,--\,24\,700\,\AA{}.

\subsection{Telluric line removal}
\begin{figure*}
\centering
\includegraphics[width=8.5cm]{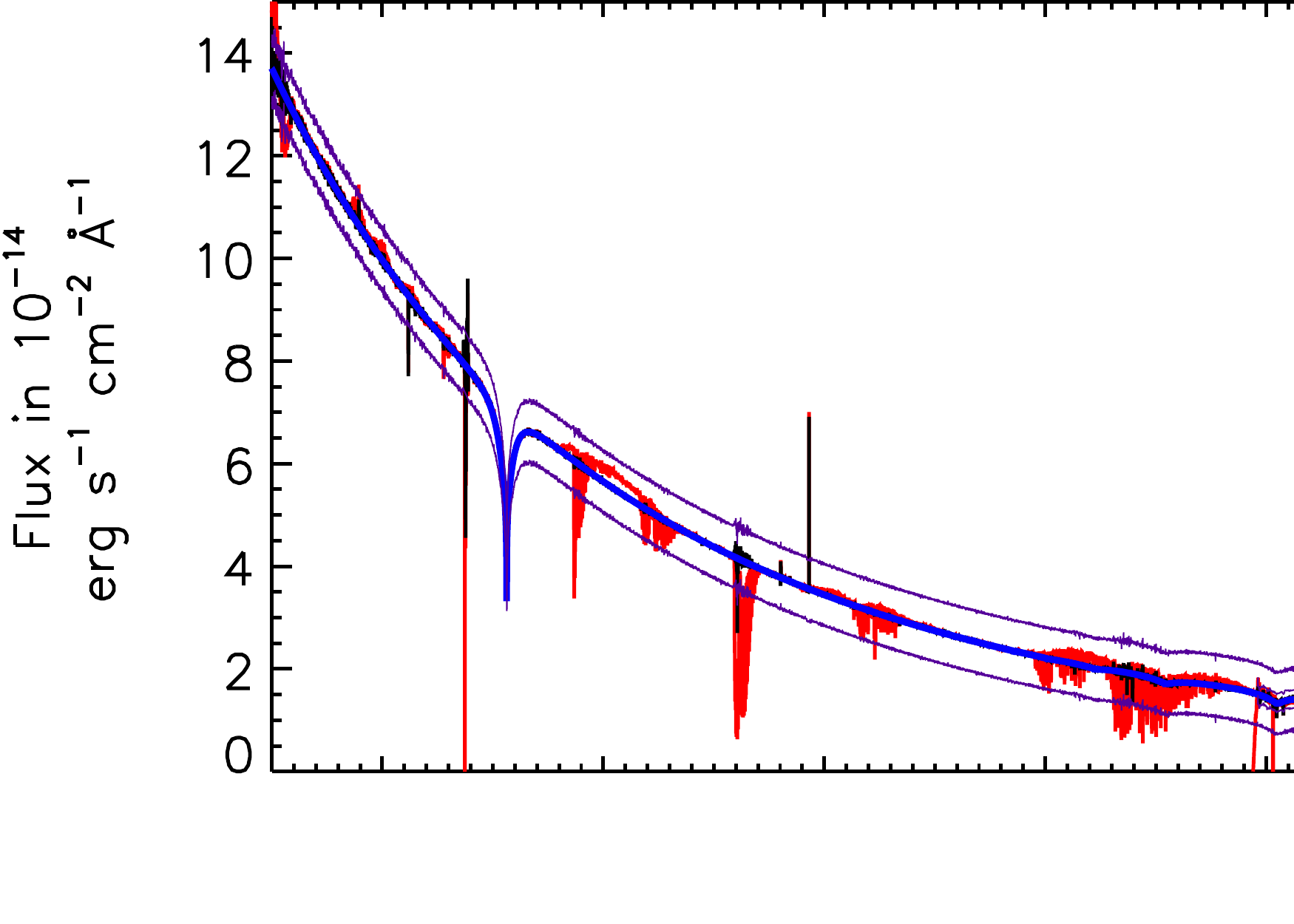}\vspace{-2mm}\includegraphics[width=8.5cm]{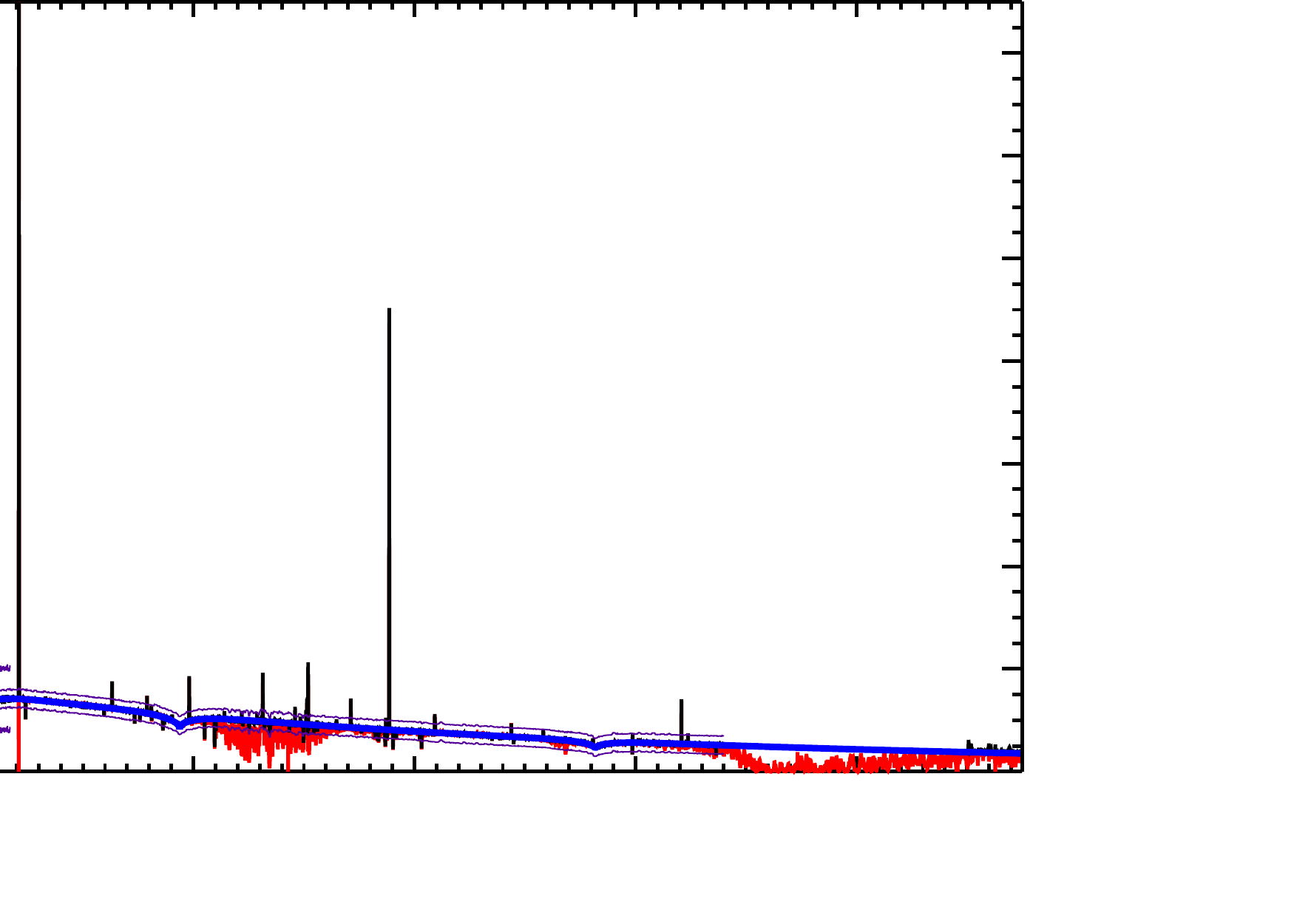}\\[-8.5mm]
\includegraphics[width=8.5cm]{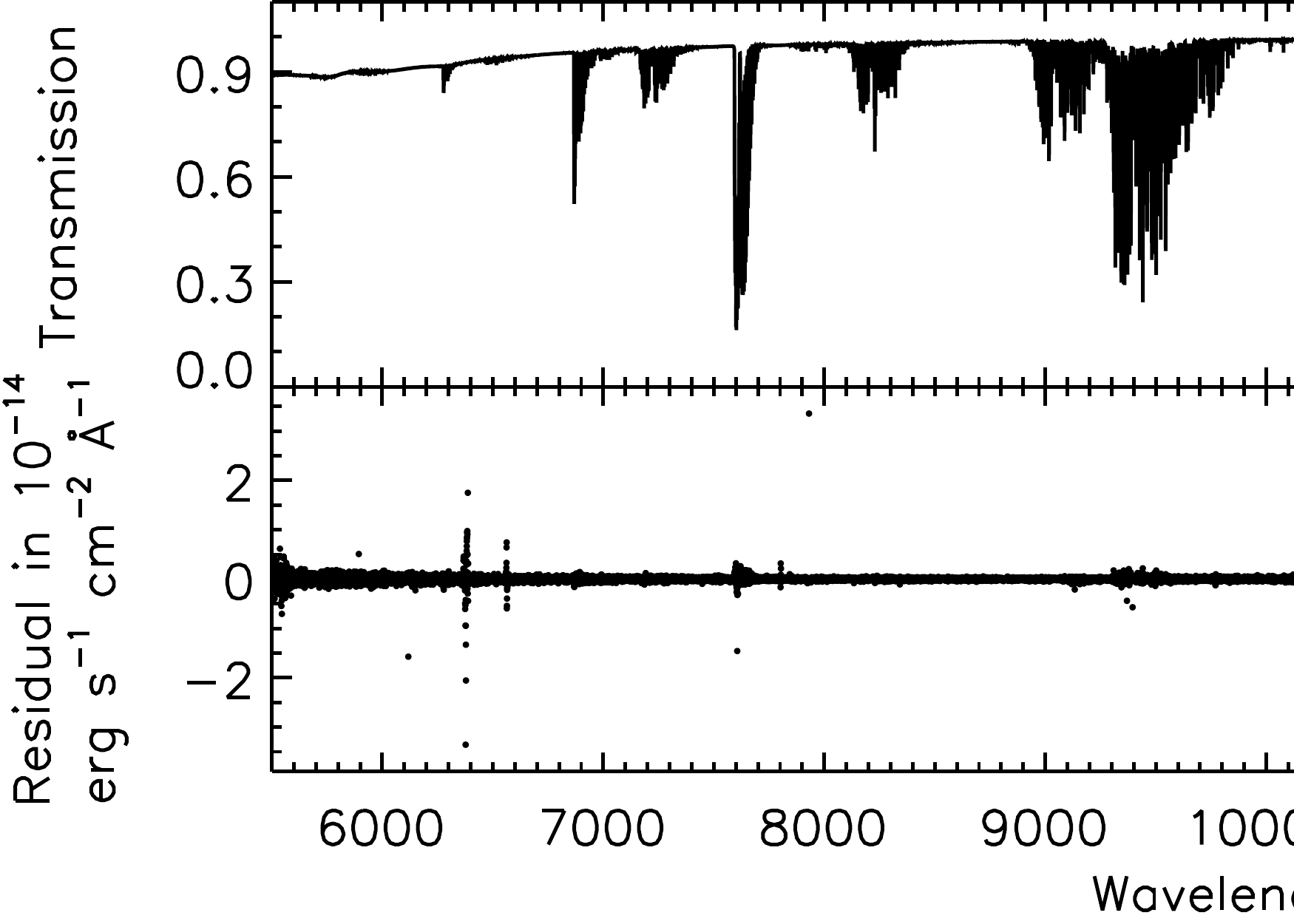}\vspace{-2mm}\includegraphics[width=8.5cm]{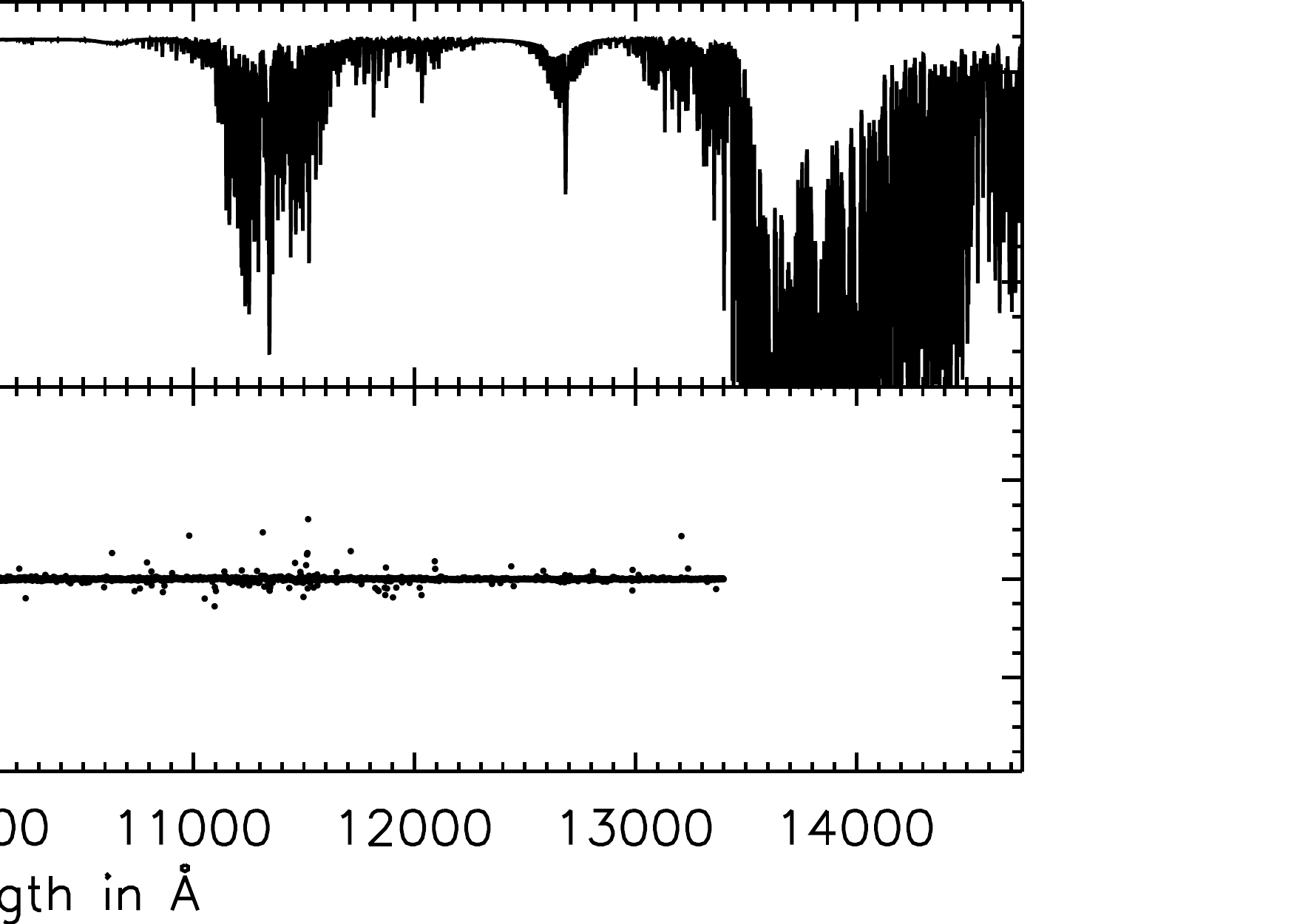}\\[2mm]
\includegraphics[width=8.5cm]{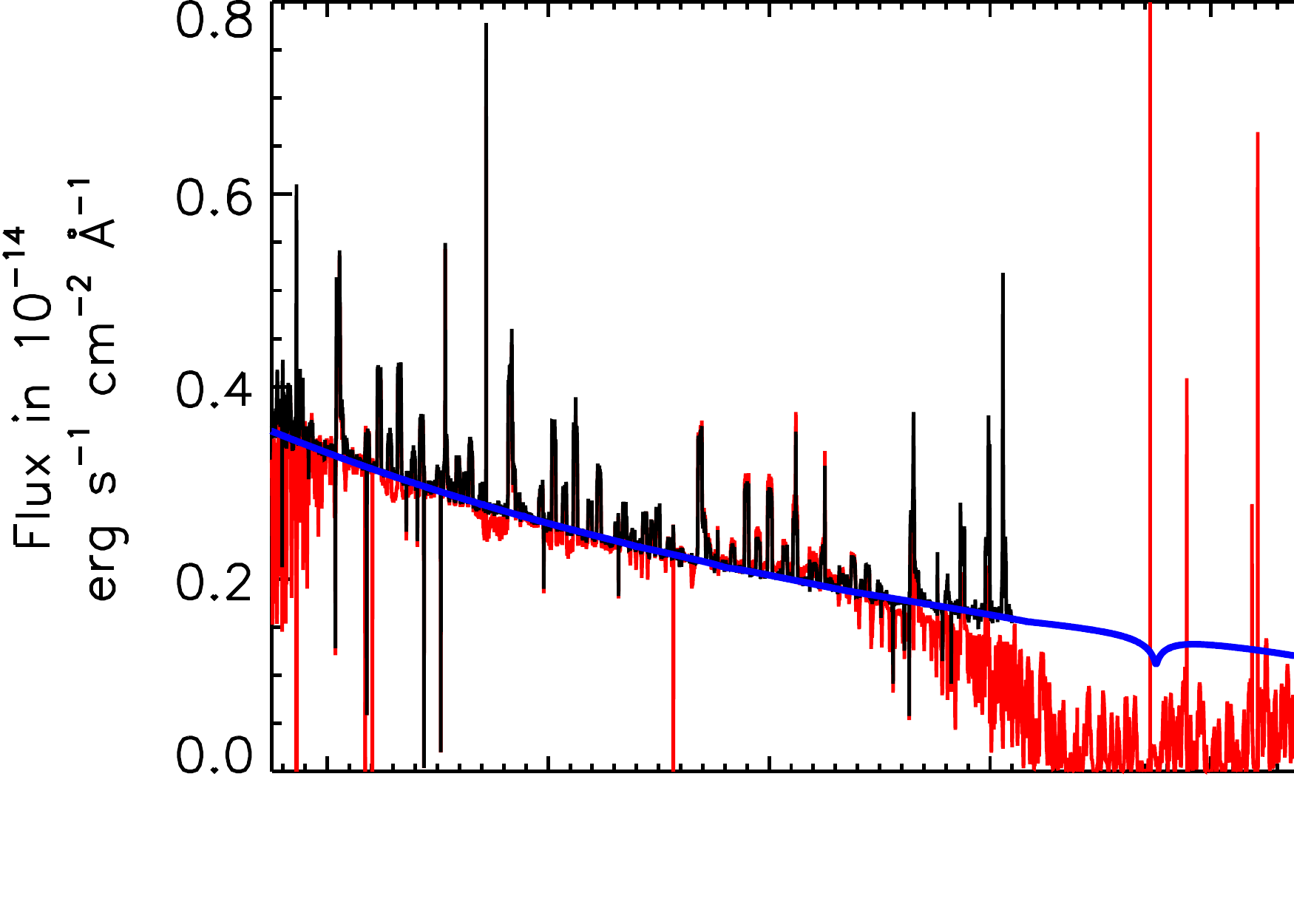}\vspace{-2mm}\includegraphics[width=8.5cm]{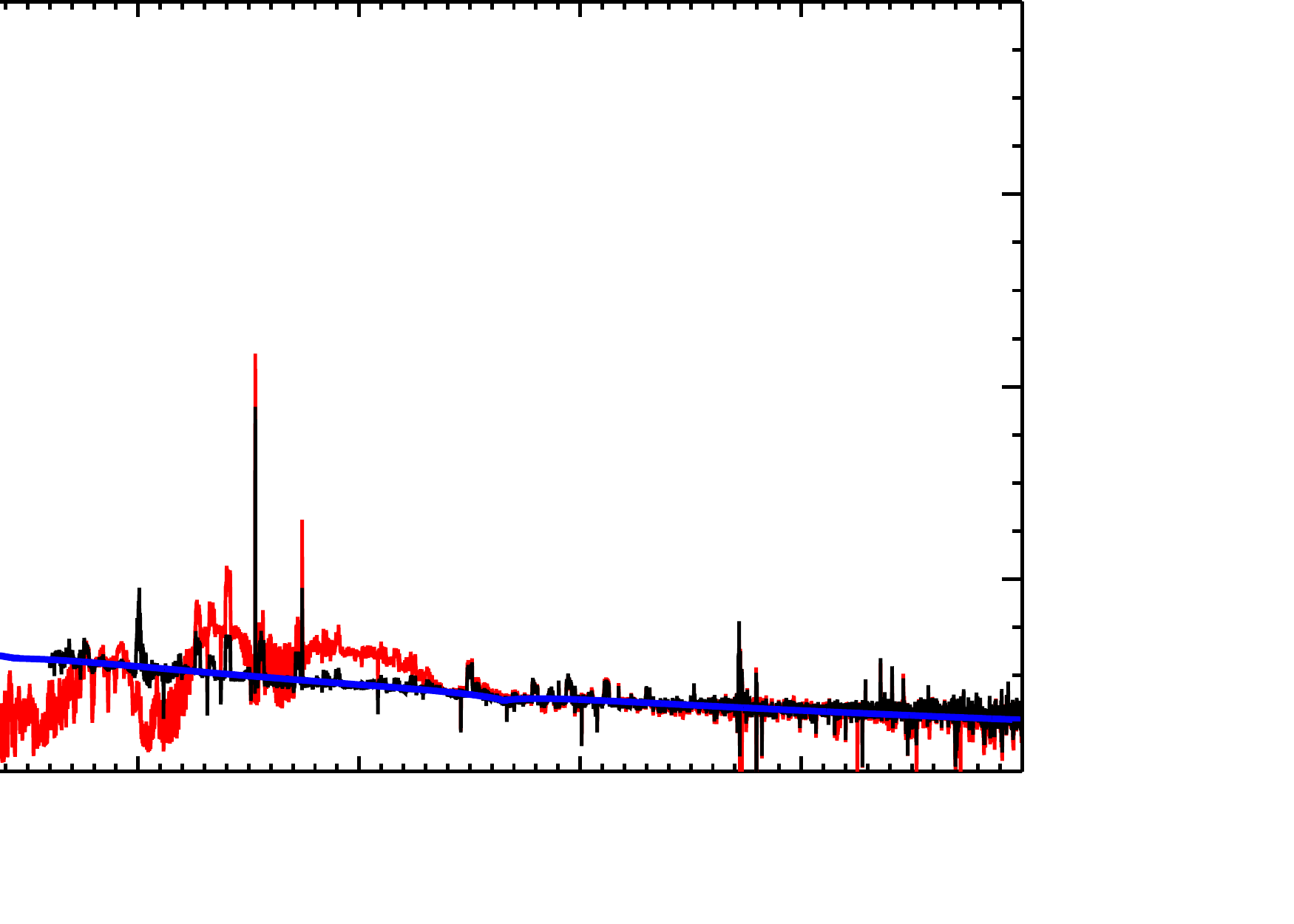}\\[-8.5mm]
\includegraphics[width=8.5cm]{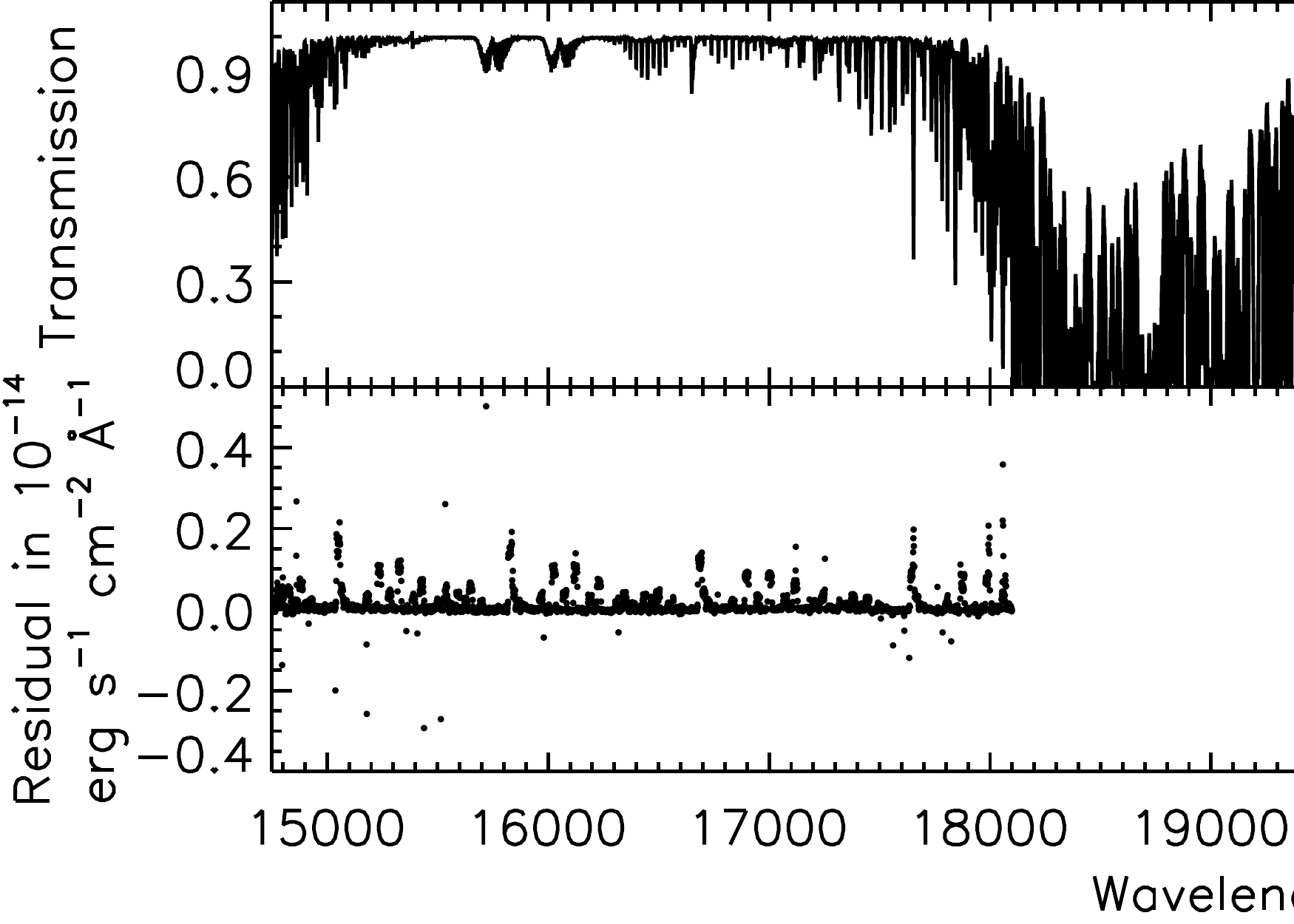}\vspace{-2mm}\includegraphics[width=8.5cm]{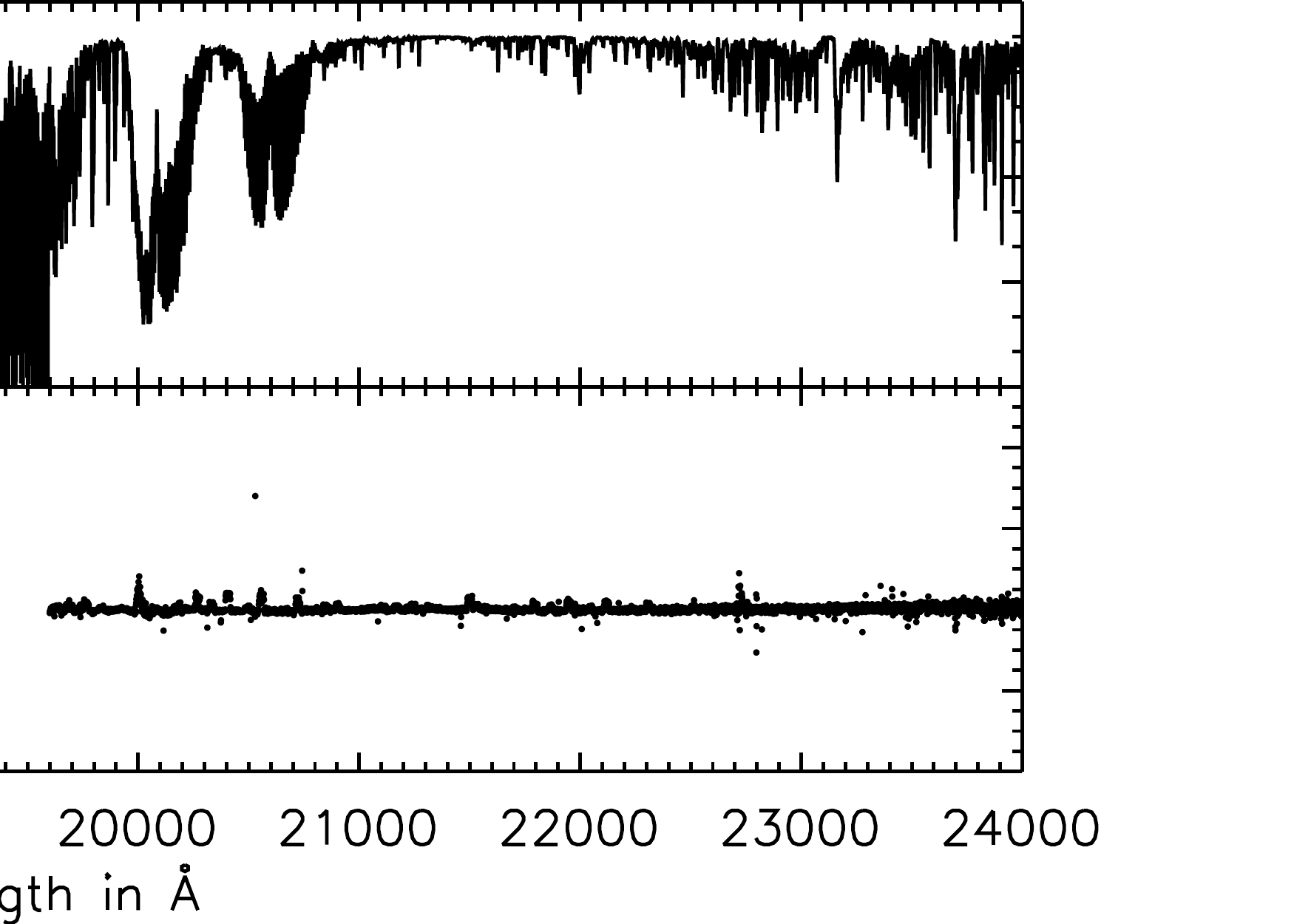}\\[2mm]
\caption{Telluric line corrected spectrum of the flux standard white dwarf EG 274 (\textit{black}), pipeline flux-calibrated spectrum (\textit{red}) for comparison, catalogued flux (\textit{blue}), and limits to identify detector artefacts (\textit{thin purple}) in the upper panel, the telluric transmission model in the middle panel, and the residuals (telluric corrected minus catalogued flux) in the bottom panel of the plots. Residuals are typically a small percentage of the actual flux value. Typical statistical uncertainties of the residuals are smaller than the plot symbols. We do not include the uncertainty of the absolute flux calibration here. Extreme residuals due to detector artefacts fall outside the displayed range. See text for further comments on the construction of this figure.}
\label{fig:completespecwm}
\end{figure*}

The performance of the telluric line removal method is illustrated in Fig.~\ref{fig:completespecwm}. Here, the corrected VIS (upper panel) and NIR (lower panel) arm spectra of EG~274 are shown in black together with the pipeline flux-calibrated spectrum in red and the catalogued flux in blue; the lower parts of the plot show the modelled telluric transmission spectra and the residuals, i.e. the telluric corrected spectrum minus the catalogued flux.
We derived the detector response curve for the white dwarf EG~274 using the same spectrum as shown here. Thus, the overall agreement, i.e. on scales larger than the size of the box-car used for smoothing, between the catalogue flux and the stellar flux presented is good by construction. We show Fig.~\ref{fig:completespecwm} on an absolute flux scale and not as a normalised spectrum to indicate where the spectrum is dominated by white dwarf features. On the other hand, we chose to smooth the response curve on fairly large scales (see Sect.~\ref{sec:obsdatared}) and all features smaller than this scale \emph{do} depend on our method to remove the telluric lines. Thus, Fig.~\ref{fig:completespecwm} is intended to demonstrate the performance of our method on scales much narrower than the size of the smoothing used for the response curve.

An inspection of the modelled transmission spectrum shows that in the wavelength regions between 13\,400\,\AA{} and 14\,500\,\AA{} and between  18\,100\,\AA{} and 19\,600\,\AA{} the transmitted flux is essentially zero, and therefore the true flux cannot be reconstructed. For most of the spectrum no residuals of telluric lines are visible, making the corrected spectrum indistinguishable from the catalogued flux. The emission-like features visible in the observed spectrum longward of 15\,000\,\AA{} are sky emission lines that were not entirely removed by the pipeline. These are also visible in the residuals. It is important to note the differences between the pipeline flux calibration (red) and our flux calibration (black), most pronounced in the region between 20\,300\,\AA{} and 21\,400\,\AA{}. These differences are mostly due to the interpolation performed by the pipeline in regions with telluric lines. As we remove the telluric lines before we determine the response, we can use these regions and accurately follow the response over the entire spectral region reaching a better flux calibration than the ESO pipeline. 

\begin{figure*}
\centering
\includegraphics[width=8.5cm]{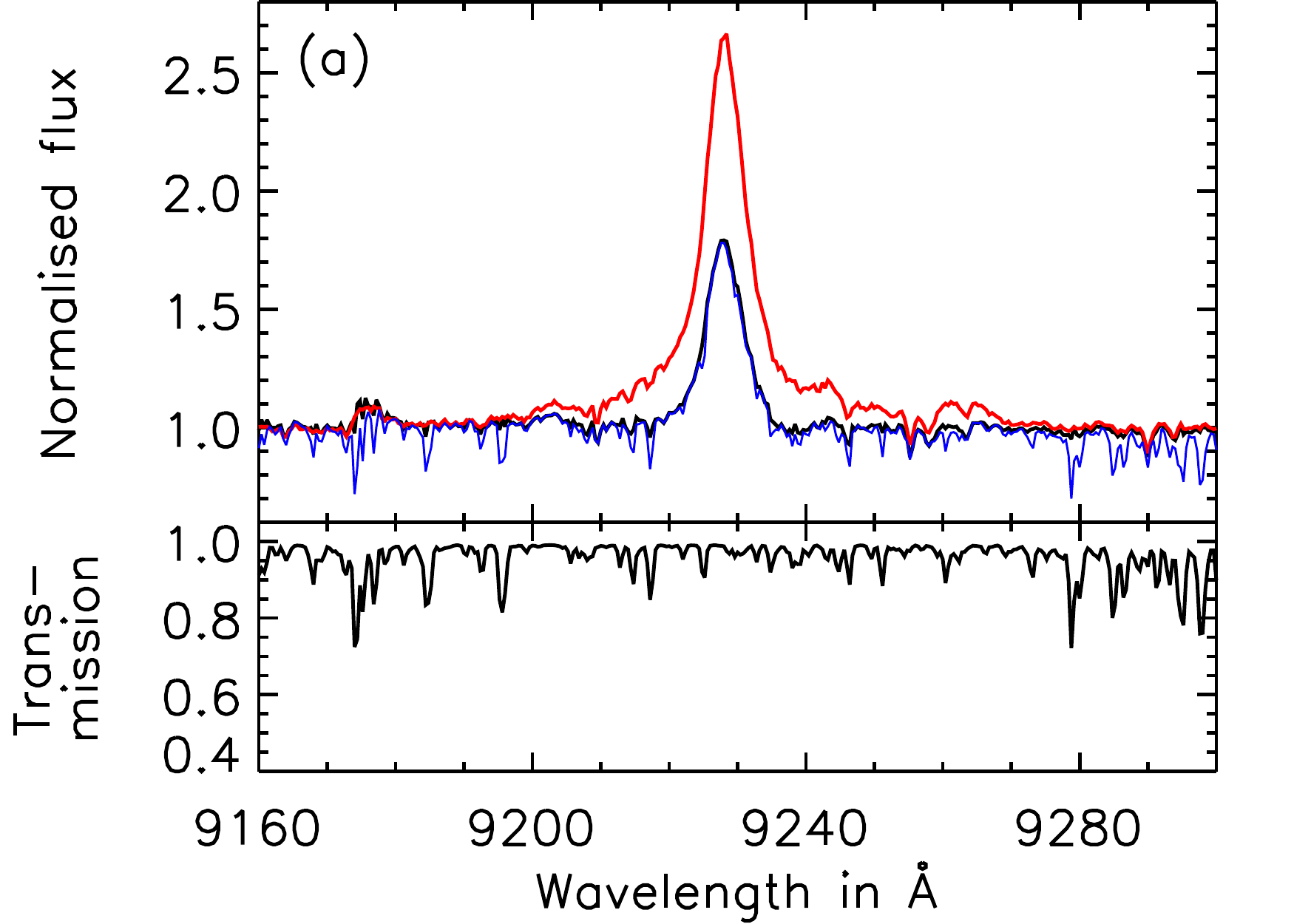}
\includegraphics[width=8.5cm]{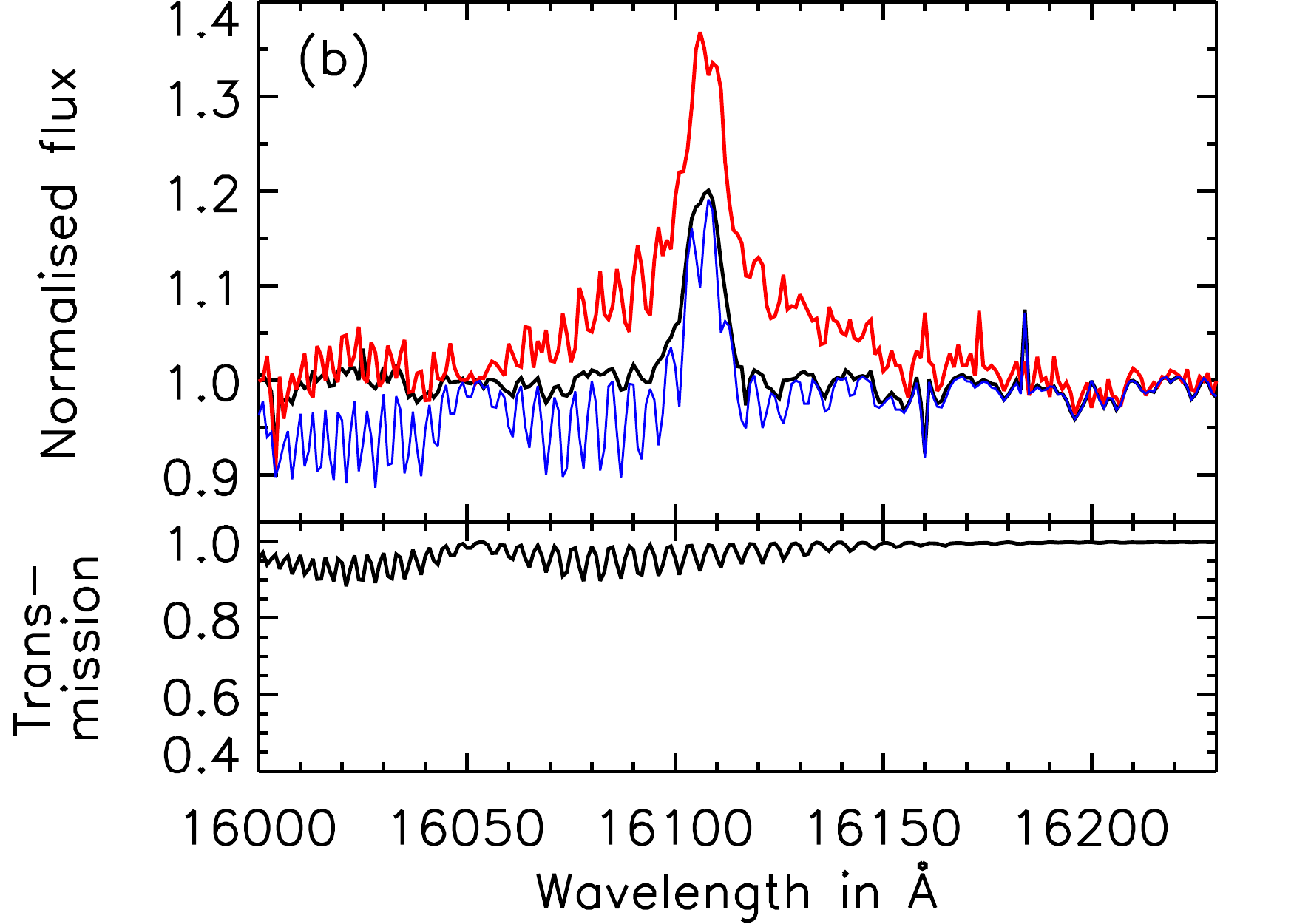}
\includegraphics[width=8.5cm]{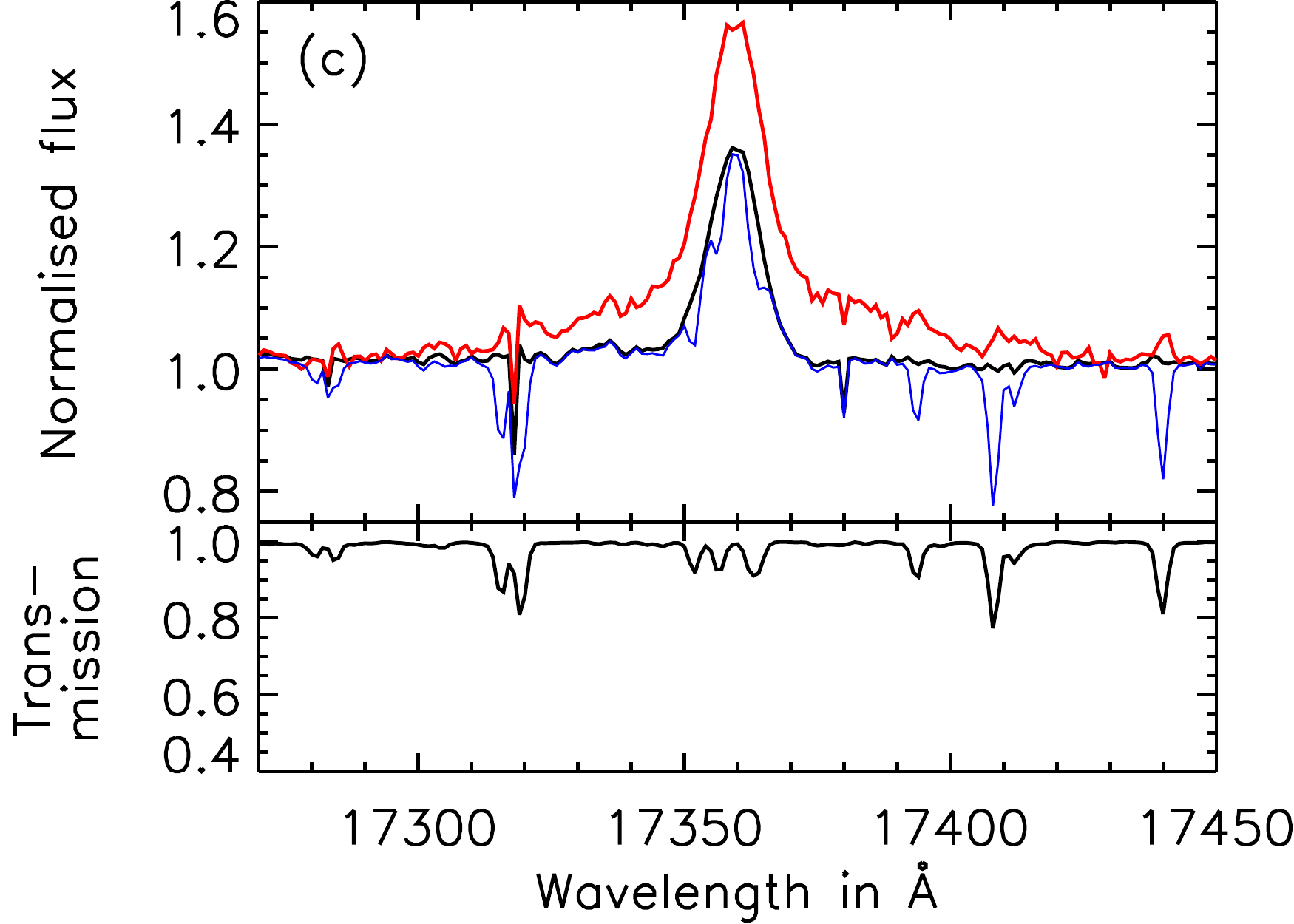}
\includegraphics[width=8.5cm]{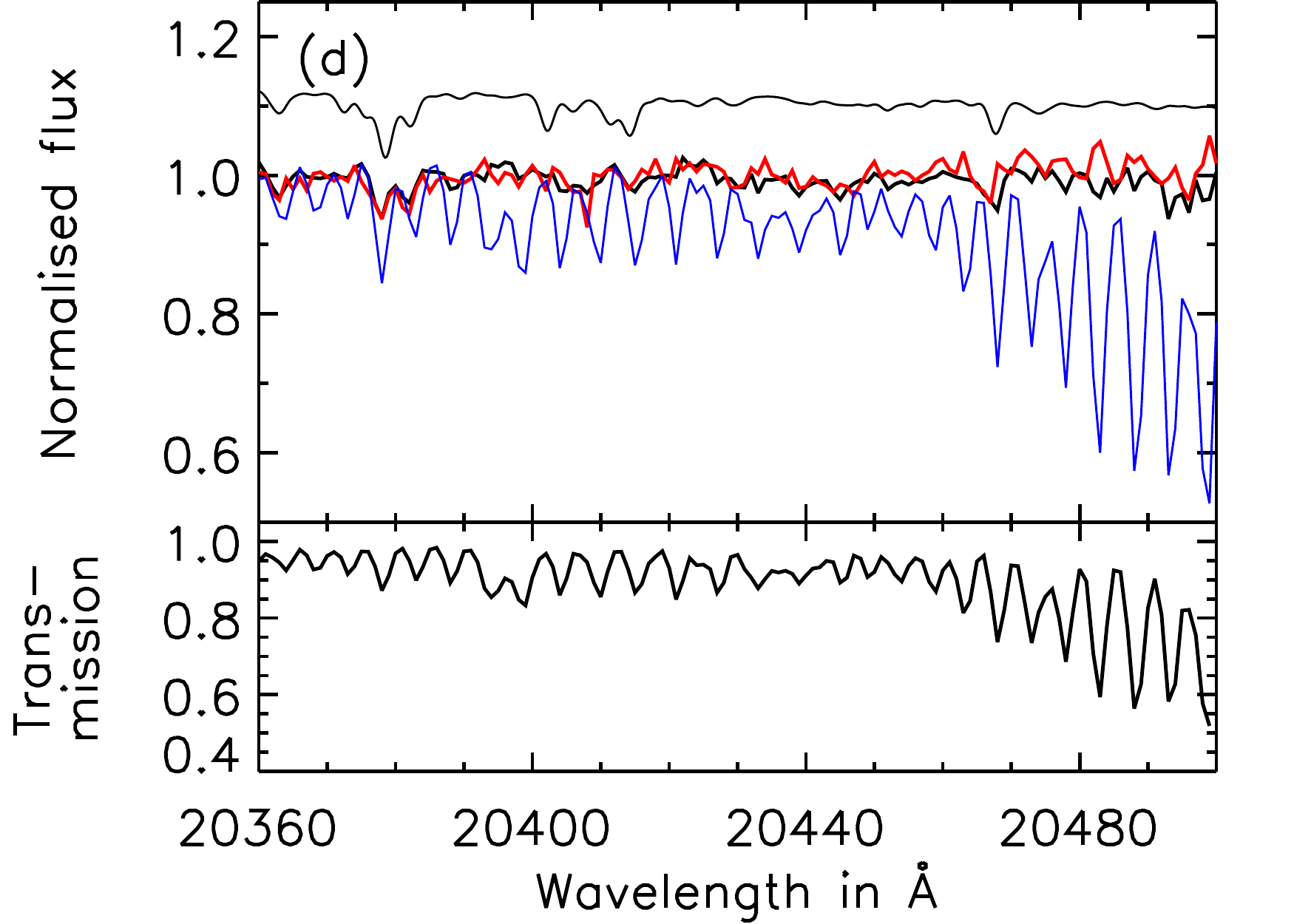}
\caption{Examples of telluric line removal leaving no discernible systematic residuals. This statement is quantified in Section~\ref{ssec:errorest}. Corrected spectrum (\textit{black}) and for comparison the original spectrum (\textit{thin blue}), as well as the spectrum corrected using IRAF (\textit{red}) in the upper panels normalised for better comparability, as well as the corresponding telluric model in the lower panels of the individual plots: hydrogen lines \textit{(a)} Paschen 9 of CTTS V895 Sco, \textit{(b)} Brackett 13, and \textit{(c)} 10 of CTTS S CrA, as well as \textit{(d)} continuum region of WTTS MV~Lup. In the upper panel of subplot \textit{(d)}  a PHOENIX model for $T_{\textnormal{eff}}=4800$\,K (\textit{thin black}) is also shown (shifted up by 0.1 for clarity).}
\label{fig:goodexamples}
\end{figure*}

The  method works successfully both in continuum regions and at emission lines. Figure~\ref{fig:goodexamples} shows some examples of the corrected spectrum and the original spectrum. The spectra were normalised for better comparability. While Figs.~\ref{fig:goodexamples}\,(a) to (c) show various hydrogen emission lines of CTTS, Fig.~\ref{fig:goodexamples}\,(d) shows a continuum segment of the weak-lined T Tauri star (WTTS) MV~Lup. For the continuum segment we also show a PHOENIX model from the G\"ottingen Spectral Library by PHOENIX \citep{phoenixmodels} for $T_{\textnormal{eff}}=4800$\,K, the effective temperature of MV~Lup \citep{1997A&A...320..185W,2003A&A...404..913S}, to illustrate the intrinsic absorption lines of the star.

\begin{figure*}
\centering
\includegraphics[width=8.5cm]{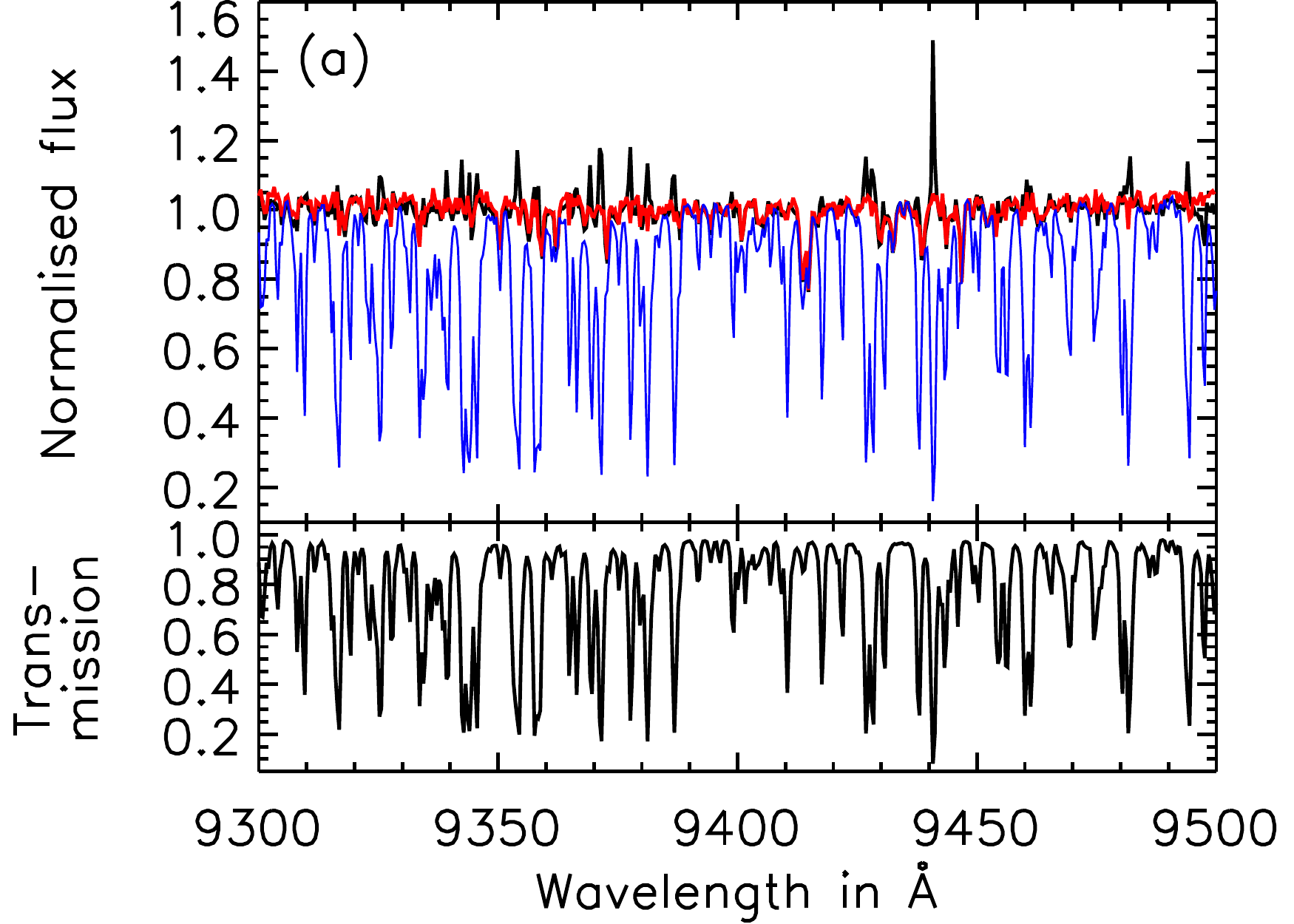}
\includegraphics[width=8.5cm]{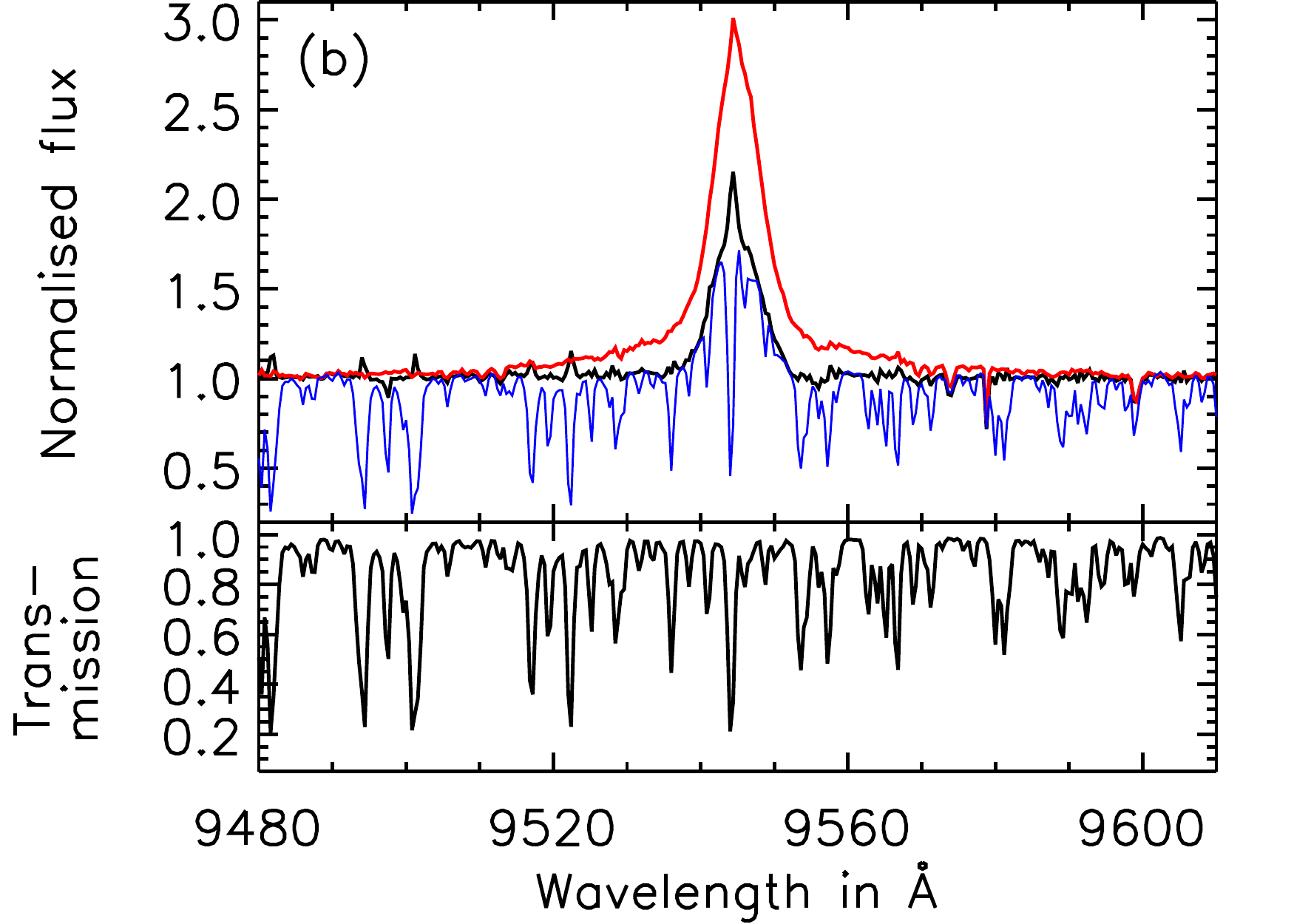}
\caption{Examples of telluric line removal leaving residuals. Corrected spectrum (\textit{black}) and for comparison the original spectrum (\textit{thin blue}), as well as the spectrum corrected using IRAF (\textit{red}) in upper panels, normalised for better comparability, as well as the corresponding telluric model in the lower panels of the individual plots: \textit{(a)} continuum region of WTTS MV Lup and \textit{(b)} hydrogen line Paschen 8 of CTTS V895~Sco.}
\label{fig:badexamples}
\end{figure*}

\citet{2010A&A...524A..11S} point out that using telluric transmission models to remove telluric lines is mainly limited by the accuracy of the information in the HITRAN database, especially for water lines, which they say are ``the hardest to model''. Not surprisingly,  our method also fails to provide a good fit of the telluric lines in some spectral ranges. These failures are illustrated in the following examples.
In Fig.~\ref{fig:badexamples} we show the region around 9400\,\AA{}, which is dominated by water absorption lines. Both in the continuum segment of the WTTS MV~Lup (Fig.~\ref{fig:badexamples}\,(a)) and in the hydrogen emission line Paschen 8 of CTTS V895~Sco (Fig.~\ref{fig:badexamples}\,(b)) there are obvious residuals from poorly modelled  water lines. We tried to obtain a better fit of the telluric lines, for example by narrowing the fitted region, but did not succeed. We conclude that inaccuracies in the HITRAN database are the likely cause for our failure to reach a better fit.

Apart from the few segments where the telluric line removal cannot be performed or is unsuccessful for the reasons stated above, in general the telluric lines are removed quite successfully.
\begin{figure*}
\centering
\includegraphics[width=8.5cm]{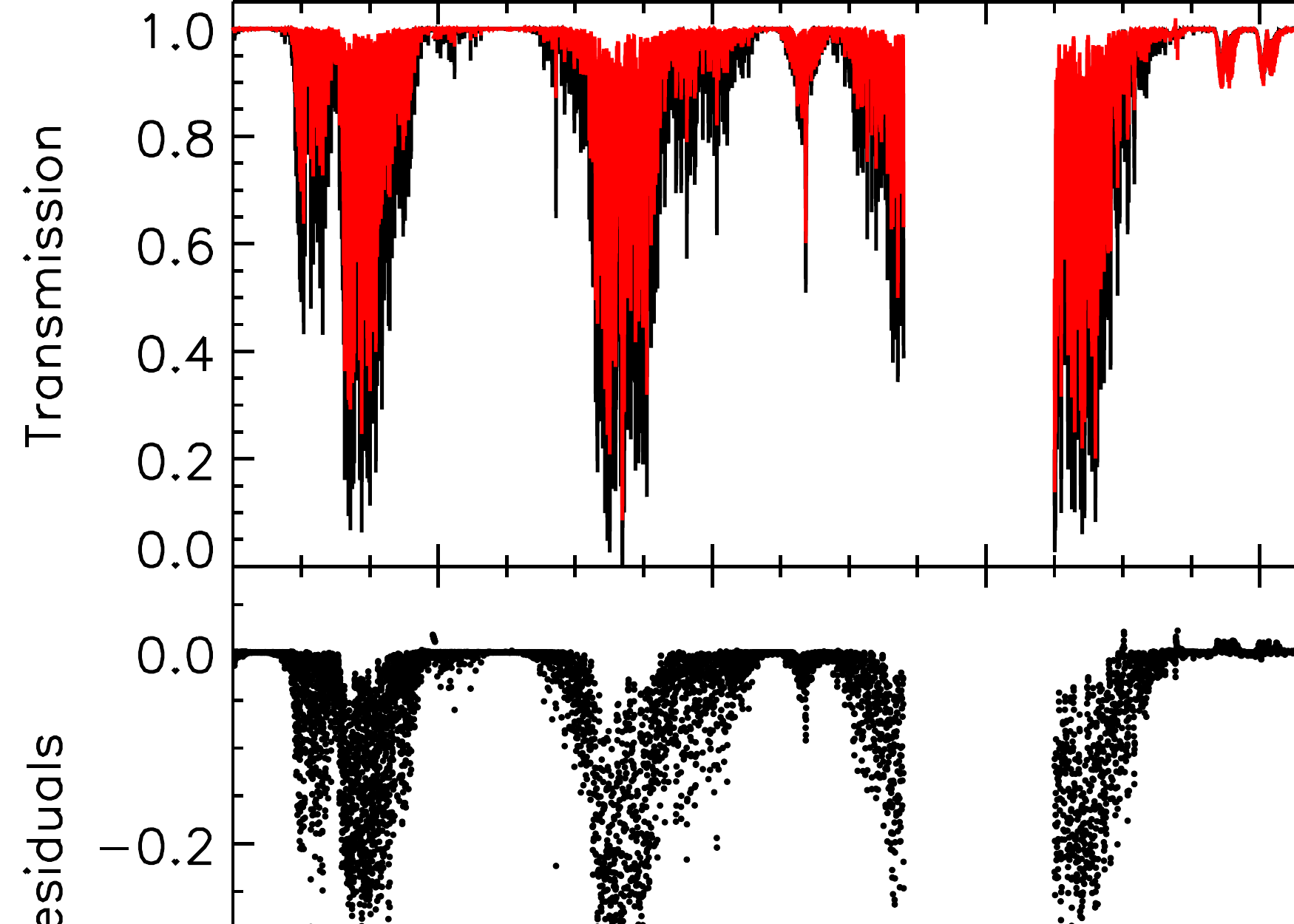}\vspace{-2mm}\includegraphics[width=8.5cm]{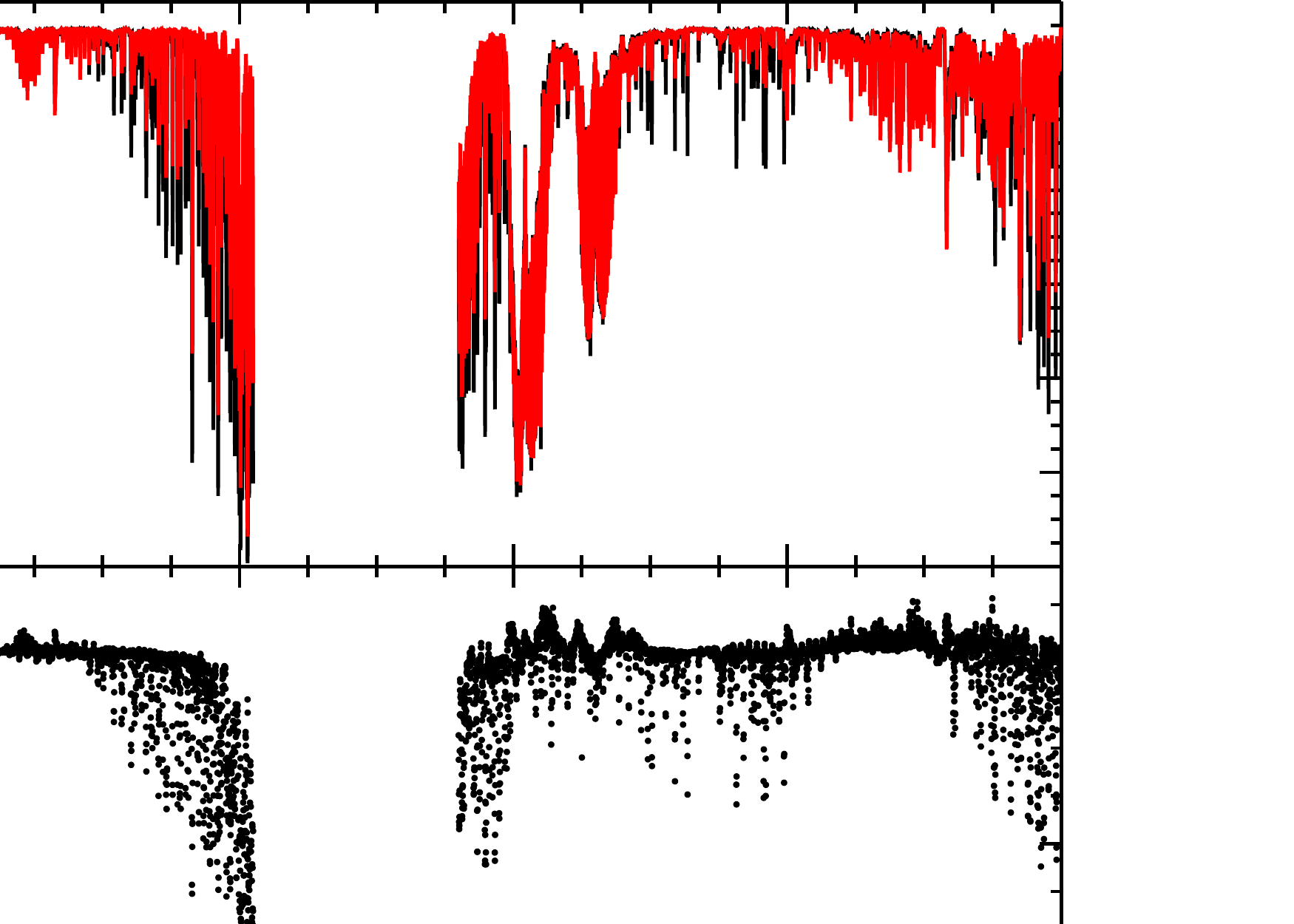}\\[0.16cm]
\includegraphics[width=8.5cm]{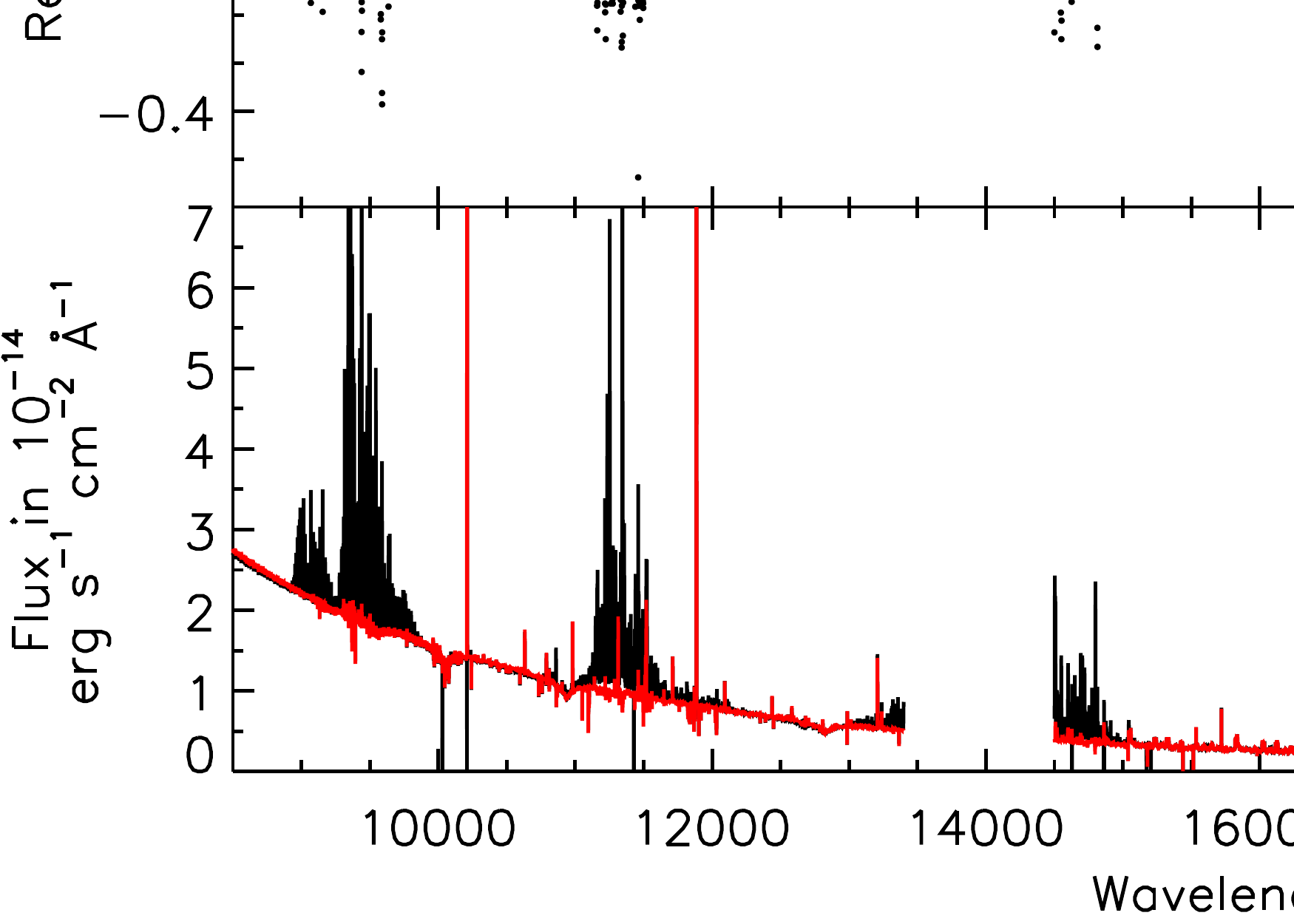}\vspace{-2mm}\includegraphics[width=8.5cm]{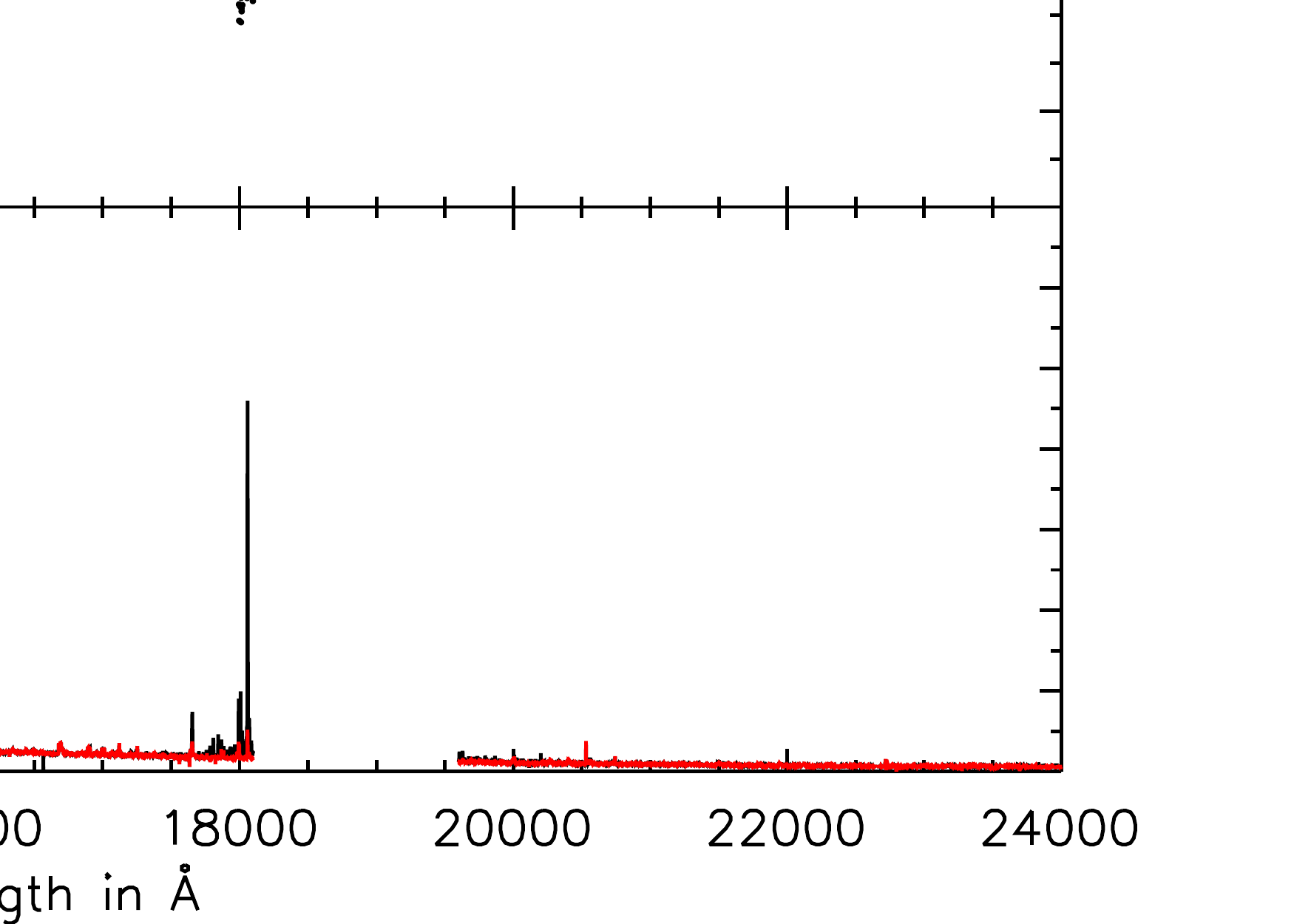}
\caption{Comparison of the telluric model calculated for EG~274 (\textit{red}) and a standard ATRAN model (\textit{black}) in the upper panel and the residuals (ATRAN model minus EG~274 model) in the middle panel. The bottom panel shows the spectrum of EG~274 corrected with the standard ATRAN model (\textit{black}) and corrected with the telluric model specifically calculated for it (\textit{red}).}
\label{fig:comp_atran}
\end{figure*}

\subsection{Comparison with the IRAF task \texttt{telluric}}
As already described in the Introduction (Sect.~\ref{sec:intro}), the typical procedure used to remove telluric absorption employs the observation of a standard star. The IRAF task \texttt{telluric} is widely used for this purpose. We performed the telluric line removal for a few objects from the sample with this software and the telluric standard stars (all of which are of spectral type B) taken during the observation run to compare the results to our modelling approach. In Figs.~\ref{fig:goodexamples} and \ref{fig:badexamples} the spectrum corrected using IRAF, normalised for better comparability, is shown in red. In regions where the standard star has no lines, the results from both methods are quite similar (e.g.\ Fig.~\ref{fig:goodexamples}\,(d)). In the above-mentioned region dominated by water absorption  the correction with IRAF is better than the one using our method (Fig.~\ref{fig:badexamples}\,(a)). The IRAF task \texttt{telluric} does not take into account the intrinsic hydrogen absorption lines of the standard star. Thus, when using the standard star spectrum to correct for telluric lines in the regions of hydrogen lines, these intrinsic lines strongly distort the object spectrum  (see Figs.~\ref{fig:goodexamples}\,(a) to (c) and \ref{fig:badexamples}\,(b)). Since we directly model the transmission, correcting the spectrum in these regions poses no difficulty for our method and the line shape of the emission lines can be reconstructed.

\subsection{Comparison of telluric model with standard ATRAN model}
We also compared the telluric transmission model derived for the white dwarf EG~274 to a telluric transmission spectrum generated by the software tool ATRAN \citep{atran} in the version provided by the SOFIA Science Center\footnote{https://atran.sofia.usra.edu/cgi-bin/atran/atran.cgi}. We ran ATRAN with standard input parameters for the observation of EG~274 without any specific water vapour overburden. The parameters used following the recommendations of the SOFIA Science Center are summarised in Table~\ref{tab:atran_params}; for more details see the explanations by the SOFIA Science Center and \citet{atran}. Since ATRAN produces an absorption spectrum without taking into account continuum effects, we reran LBLRTM in its ``no continuum'' mode using the previously determined parameters for EG~274. Both spectra were convolved with Gaussians according to the previously determined parameters. Figure~\ref{fig:comp_atran} shows both spectra and the difference between them, i.e. the ATRAN spectrum minus the LBLRTM spectrum. The ATRAN model clearly underpredicts the depth of the lines in many cases. The bottom panel of Fig.~\ref{fig:comp_atran} also shows the spectrum of EG~274 corrected for telluric absorption using both model spectra. For a good correction the telluric spectrum has to be modelled as closely as possible. It is thus important to adapt the model to the current conditions in the atmosphere.
\begin{table}
\caption{Input parameters used for the ATRAN model.}
\label{tab:atran_params}
\centering 
\begin{tabular}{ll}
\hline\hline
Parameter & Value\\
\hline
Observatory altitude & 8688\,ft\\
Observatory latitude & 39\,deg\\
Water vapour overburden & 0\\
Number of atmospheric layers & 2\\
Zenith angle & 38.4\,deg\\
Resolution & 0\tablefootmark{a}\\
\hline
\end{tabular}
\tablefoot{\tablefoottext{a}{No smoothing to a specific resolution is performed.}}
\end{table}
\subsection{Error estimation}\label{ssec:errorest}
In order to evaluate the performance of our method in a statistical sense we use the spectrum of the white dwarf EG~274. We use the catalogued flux $F_{\mathrm{cat}}$ as the true model and the observed telluric line corrected flux $F^{\mathrm{t.c.}}_{\mathrm{obs}}$ as ``data'' along with the statistical errors of the observed telluric line corrected flux $\sigma^{\mathrm{t.c.}}_{\mathrm{obs},\,i}$ (these are simply the errors calculated by the data reduction pipeline scaled by the telluric correction).

The data contains several detector artefacts, most of them in the NIR arm, for example\ at 10\,200\,\AA{} or 11\,900\,\AA{}. Since these artefacts are unrelated to the method that we want to test in this section, we exclude affected data points from the analysis presented here. Additionally, we restrict our analysis to the VIS arm and the first part of the NIR arm (9900\,--\,13\,400\,\AA{}, called NIR1 in the following) because the residuals of the sky lines already cause a deviation from the catalogued flux in the second and third part of the NIR arm that would be mixed with the residuals of the telluric line removal.

We determine affected data points in the following way. We compare a median filtered version of the data (three bins in the VIS arm, seven bins in NIR1) to the original data and exclude all data points in the original data that are more than $6\cdot10^{-15}$\,erg\,s$^{-1}$\,cm$^{-2}$\,\AA{}$^{-1}$ (VIS) or $1.75\cdot10^{-15}$\,erg\,s$^{-1}$\,cm$^{-2}$\,\AA{}$^{-1}$ (NIR1) away from the median filtered version (see purple line in Fig.~\ref{fig:completespecwm}). In the VIS arm we additionally exclude a larger artefact between 6366\,\AA{} and 6390\,\AA{} by hand. We exclude 101 of the 11675 data points in the VIS arm and 47 of the 3440 data points in NIR1. The statistical errors $\sigma^{\mathrm{t.c.}}_{\mathrm{obs},\,i}$ of NIR1 contain several outliers with errors up to three magnitudes larger than the average errors. Although these errors are the result of the pipeline error propagation, they seem greatly overestimated and we replace these by the mean of the neighbouring errors. This is done for 50 data points.

Because we constructed the data to match the flux of the model spectrum (see above) on large scales, uncertainties in the model or the global flux do not influence the value of $F^{\mathrm{t.c.}}_{\mathrm{obs}} - F_{\mathrm{cat}}$ we calculate here, thus EG~274 offers a clean way to characterise the errors introduced by our telluric line removal method.

First of all, we look at the $\chi^2$ test statistic. Regions unaffected by telluric lines give an estimation of the data quality independent of the telluric line removal. We find the following reduced $\chi^2$ values: $\chi^2_{\mathrm{unaff,\,VIS}}=1.23$ for the VIS arm and $\chi^2_{\mathrm{unaff,\,NIR1}}=2.49$ for NIR1. The pipeline errors seem to be underestimated, especially in the NIR1. In regions affected by telluric lines we find $\chi^2_{\mathrm{aff,\,VIS}}=2.14$ and $\chi^2_{\mathrm{aff,\,NIR1}}=5.67$. To reach a similar $\chi^2$ in the affected  and the unaffected regions the errors need to be increased by $\sim$30\,\% in the VIS arm and $\sim$50\,\% in NIR1.

\begin{figure}
 \includegraphics[width=\hsize]{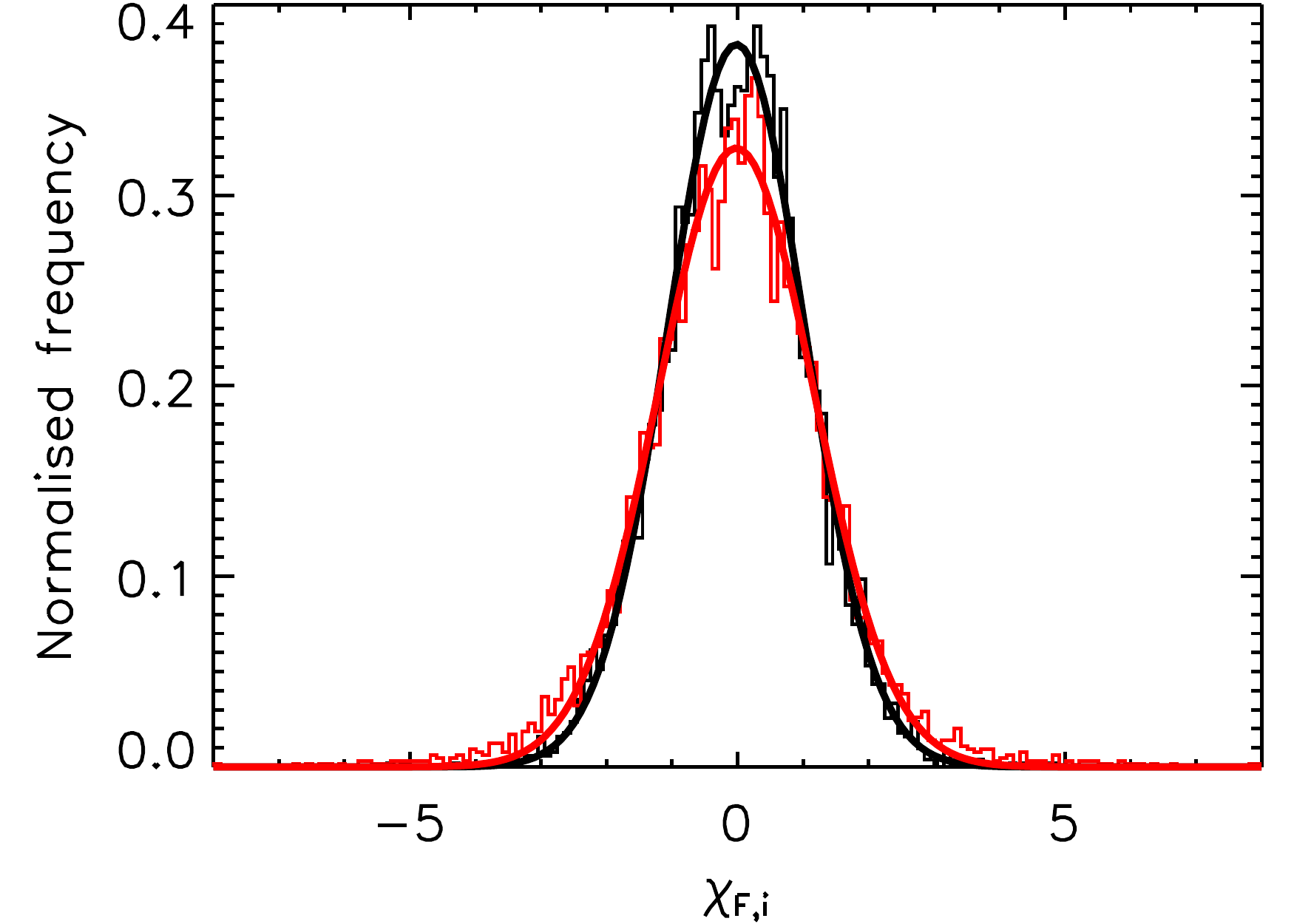}\\[3mm]
 \includegraphics[width=\hsize]{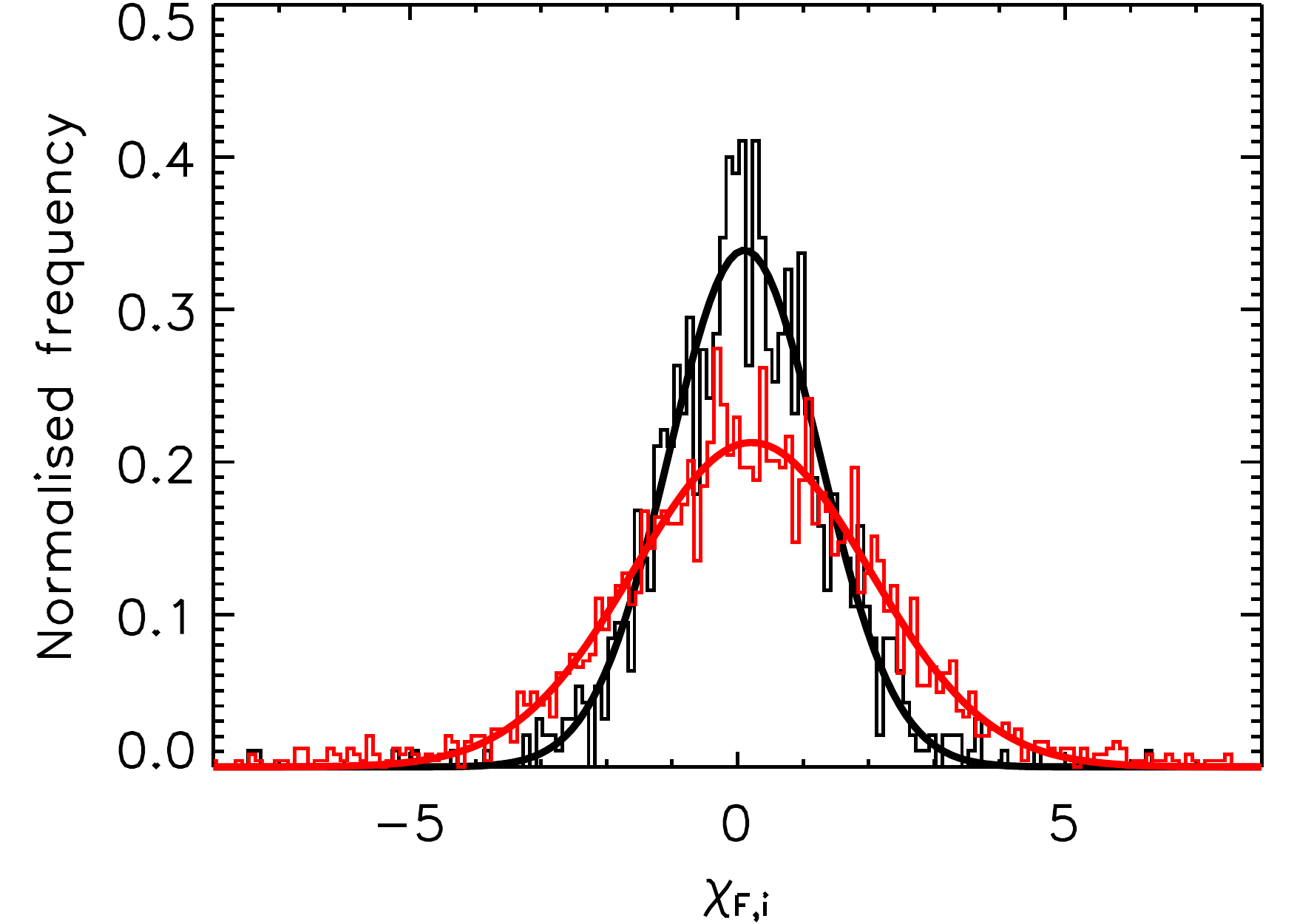}
 \caption{Distribution of $\chi^{}_{\textnormal{F},\,i}$ for VIS arm (\textit{top}) and first part of NIR arm (\textit{bottom}), both  with Gaussian fit. \textit{Black}: regions without telluric
lines in the uncorrected spectrum and \textit{red}: regions with telluric lines in the uncorrected spectrum.}
 \label{fig:chi}
\end{figure}

To check if the overall statistical distribution is well behaved, we study the distribution of the individual summands of the $\chi^2$ test statistic,
\begin{equation}\label{equ:diff}
 \chi^{}_{\textnormal{F},\,i}=\dfrac{F^{\mathrm{t.c.}}_{\mathrm{obs},\,i}-F_{\mathrm{cat},\,i}}{\sigma^{\mathrm{t.c.}}_{\mathrm{obs},\,i}}.
\end{equation}
If there were no additional error contribution by the telluric line removal, the distribution should be a Gaussian with the same standard deviation in regions without telluric lines and with telluric lines in the original spectrum. 
In the VIS arm we find a standard deviation of the Gaussian of 1.05 for the regions without telluric lines and 1.19 for the regions which are affected by telluric lines.
Thus, the standard deviation increases by 13\,\%, implying that the errors in the regions containing telluric lines are underestimated by this amount. 
In NIR1 we find standard deviations of 1.15 and 1.80, respectively. This indicates that the real errors are 113\,\% and 156\,\% of the pipeline error in the VIS arm and NIR1. We show the corresponding distributions in Fig.~\ref{fig:chi}.
The mean relative statistical error of the fluxes in each bin ($<\frac{\sigma^{\mathrm{t.c.}}_{\mathrm{obs},\,i}}{F^{\mathrm{t.c.}}_{\mathrm{obs},\,i}}>$) is 0.7\,\% in the VIS arm and 1.1\,\% in NIR1. Adding the factors derived above increases the errors to 0.8\,\% and 1.7\,\%. 

\begin{figure}
 \includegraphics[width=\hsize]{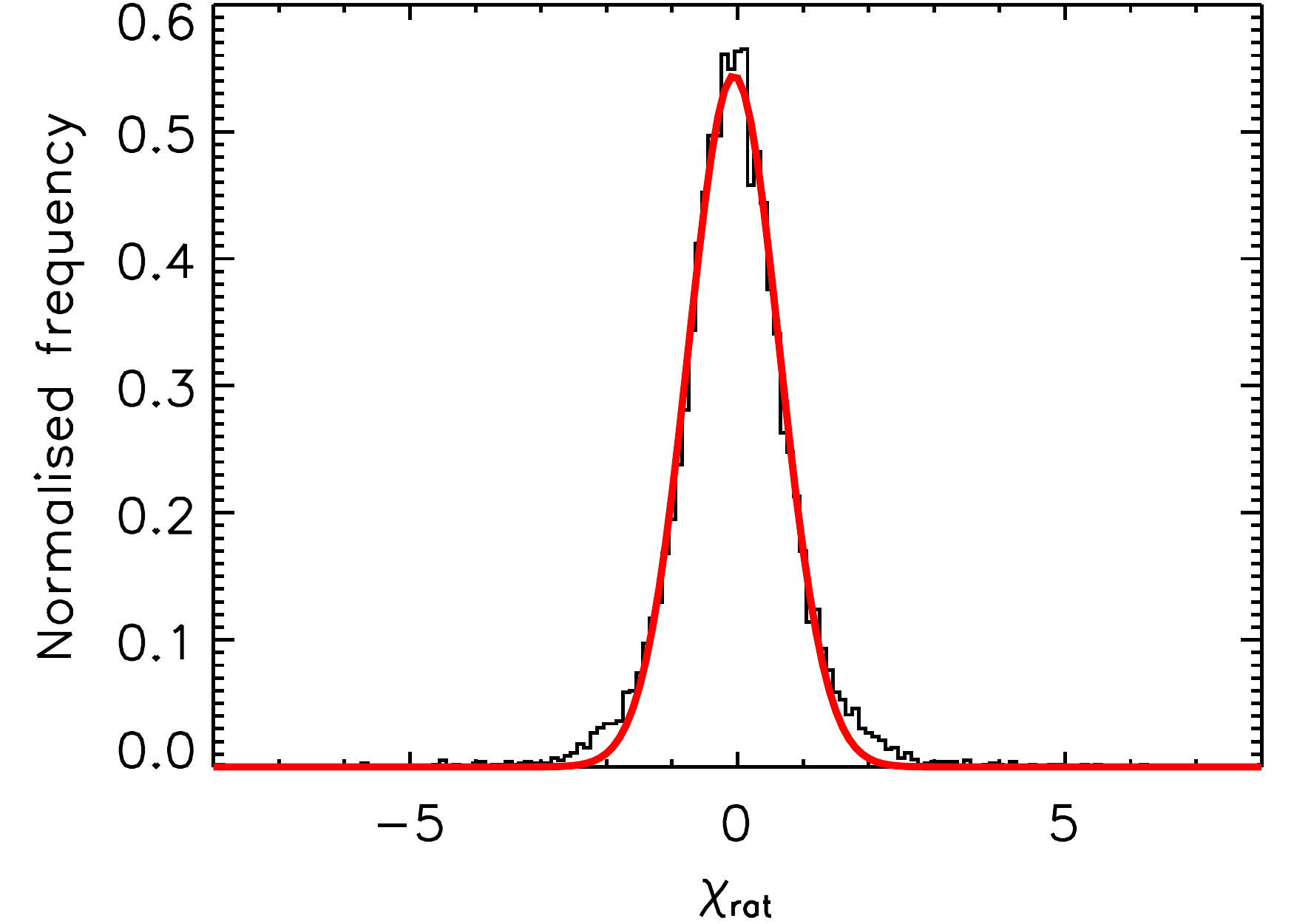}\\[3mm]
 \includegraphics[width=\hsize]{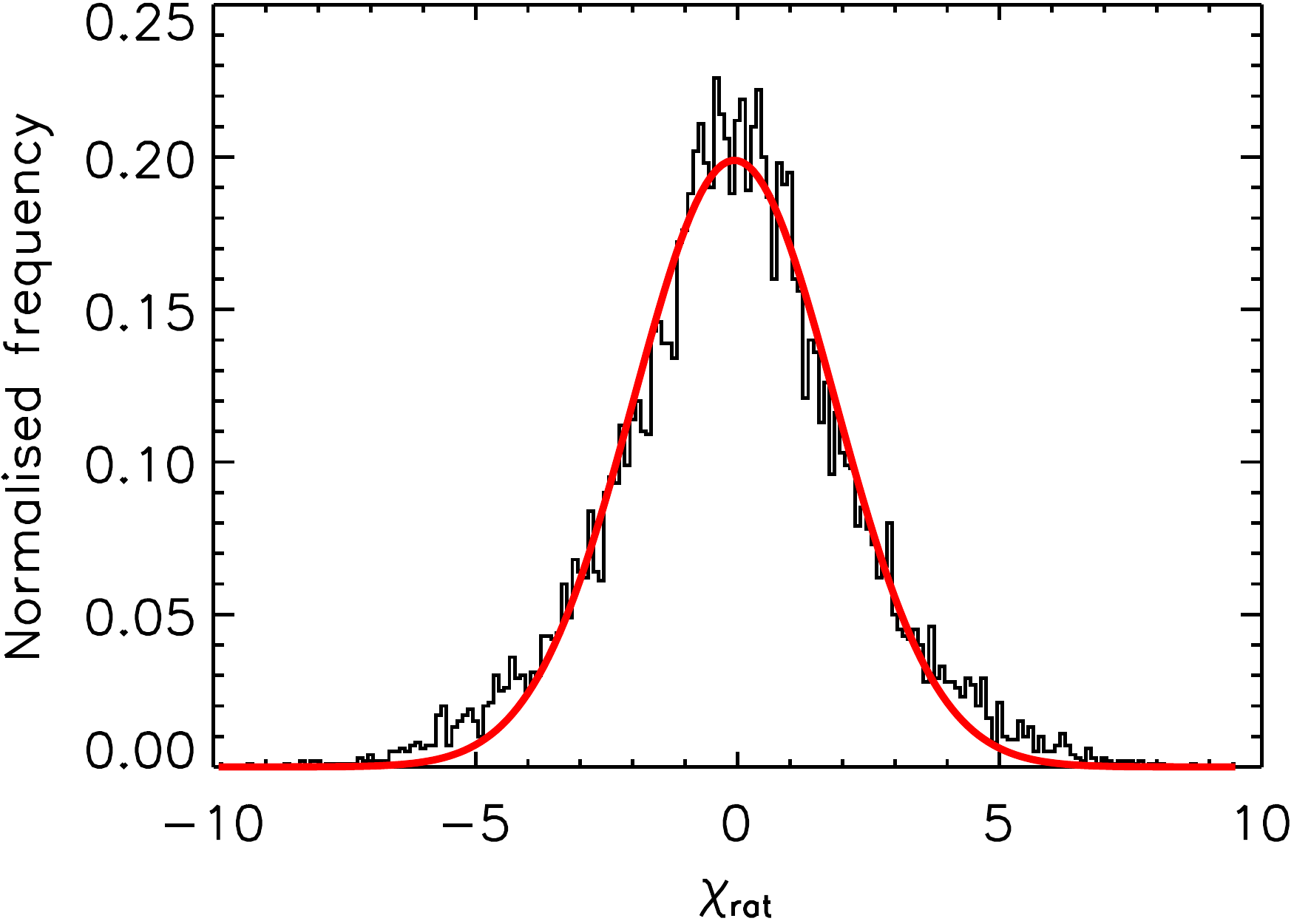}
 \caption{Distribution of $\chi^{}_{\textnormal{rat}}$ (\textit{black}) for VIS arm (\textit{top}) and first part of NIR arm (\textit{bottom}), both  with Gaussian fit (\textit{red}).}
 \label{fig:chi2}
\end{figure}

As an example of a practical application, we consider flux ratios. This allows us to check for systematic effects introduced by the telluric line removal. We randomly select from the spectrum two non-overlapping regions $a$ and $b$ of spectral width $w$, integrate the flux in the observed telluric line corrected spectrum and in the model spectrum, and calculate the flux ratio of the two regions. For the selection of the regions $a$ and $b$, we discard regions that are affected by detector artefacts as described above. We repeat this for $n$ ratios and calculate the quantity
\begin{equation}\label{equ:ratios}
 \chi^{}_{\textnormal{rat}}=\dfrac{\left(\frac{F_a}{F_b}\right)_{\mathrm{obs}}-\left(\frac{F_a}{F_b}\right)_{\mathrm{cat}}}{\sigma_{\left(\frac{F_a}{F_b}\right)_{\mathrm{obs}}}}.
\end{equation}
As pointed out above, the pipeline errors seem to be underestimated. To avoid a bias due to underestimated pipeline errors in our study, we increased the errors to 110\,\% of the pipeline error in the VIS arm and to 155\,\% in NIR1. With these error values we reach a $\chi^2\sim1$ in the regions unaffected by telluric lines. For $w=16$\,\AA{}, a typical width for hydrogen emission lines in CTTS, and $n=10\,000$, the distribution of $\chi^{}_{\textnormal{rat}}$ is approximately Gaussian. In an optimal case the standard deviation of the Gaussian should be 1. Figure~\ref{fig:chi2} shows the distributions. We find a standard deviation of 0.7 in the VIS arm and 1.9 in NIR1, indicating that the real error is only 70\% of the error that would result in $\chi^2\sim1$, which itself is already 110\,\% of the pipeline error for the VIS arm, while in NIR1 the real error is 190\,\% of the error that would result in $\chi^2\sim1$, which itself is already 155\,\% of the pipeline error. The mean relative statistical error of the $n$ ratios ($<\frac{\sigma_{\left(\frac{F_a,\,i}{F_b,\,i}\right)_{\mathrm{obs}}}}{(F_a,\,i/F_b,\,i)_{\mathrm{obs}}}>$, where the index $i$ runs over all $n$ ratios) is 0.19\,\% in the VIS arm, decreased to 0.13\,\% when subtracting the above factor. For NIR1 the mean relative statistical error is 0.67\,\%, increased to 1.31\,\%.



The numbers given here for the additional uncertainty introduced by the telluric line correction are calculated for one specific data set. We also tested the applicability of our software to lower quality data using the data set of EG~274. We created data sets with reduced resolution by increasing the bin size as well as with higher noise (the UV arm does not contain telluric features and in the other arms our spectra are binned to approximately match the spectral resolution with bin sizes $\approx0.4$~\AA{} in the VIS and $\approx1$~\AA{} in the NIR arm). For these simulated data sets, we recover the correct telluric abundances down to a bin size of 4\,\AA{} in both arms and up to noise levels increased by a factor of 10. For lower signals, the telluric abundances can no longer be fitted.
We repeated the above error estimation for the simulated data sets. For an increase in noise the relative importance of the additional error decreases. For tenfold noise the additional error is only a few percentage points. For lower resolution the results are less clear, but the general trend indicates lower additional errors for lower resolution. On the one hand, data with a lower $S/N$  leads to less precise values for the abundance of the molecular, telluric species; on the other hand, a similar \emph{absolute} contribution to the uncertainties (e.g.\ from an incomplete telluric line list) has a lower \emph{relative} influence on the uncertainties for low $S/N$ data.

\section{Summary and conclusions}\label{sec:sumcon}

We present a method to remove telluric lines from VLT/X-Shooter spectra based on the approach originally put 
forward by \citet{2010A&A...524A..11S} in the context of CRIRES. 
In addition to using the line-by-line radiative transfer code LBLRTM and a model of Earth's atmosphere to compute 
telluric transmission spectra, we utilise the very broad wavelength coverage of X-Shooter to determine the 
abundances of the most dominant producers of telluric lines (i.e. H$_{2}$O, O$_{2}$, CO$_{2}$, and CH$_{4}$) from sufficiently isolated line groups.

The computational effort to implement the telluric line modelling is manageable on a standard desktop PC. Our method removes telluric lines from both continuum regions and emission lines in general without residuals exceeding a small percentage of the initial flux except in a few isolated wavelength regions. The analysis of several error estimators shows that the method adds a systematic error to the data, which we estimate to be in the percent range. Thus, its relative importance depends on data quality and wavelength. At low quality ($S/N\sim10$) and/or shorter wavelengths the error of the corrected spectrum can be increased to $\sim$110\,\% of the statistical errors. At higher quality ($S/N\sim100$) and longer wavelengths the error can be increased to up to 190\,\% of the statistical errors. Thus, the additional statistical error introduced by our method is about 1\,\%. Improved molecular data for the telluric line simulations will hopefully improve these values in the future.

The method works successfully both on high-resolution data ($R\sim70\,000$) as used by \citet{2010A&A...524A..11S} and 
on the medium-resolution data ($R\sim10\,000$) provided by the VLT/X-Shooter instrument. Although the software package \texttt{tellrem} in its current form is tailored to handle VLT/X-Shooter spectra, its procedures can in principle be used for every instrument. The manual provides some information and aid concerning the adaption for other instruments. The software is also able to process lower quality data,  in terms of lower resolution and of higher noise.

In addition to saving precious observing time the method also circumvents the difficulties arising from the use of A or B stars as telluric standards for certain scientific applications. These stars are chosen because they have relatively featureless spectra, but they still show hydrogen lines. This complicates the situation if the specific interest is these lines in the object stars. Therefore, the use of telluric transmission models to remove telluric lines from observations is a worthwhile alternative to the observation of telluric standard stars.

\textbf{Note added in proof:} While this paper was in the refereeing process, other software tools to model the
telluric absorption have been released to the community. These alternatives are Telfit \citep{2014AJ....148...53G} and Molecfit \citep{2015A&A...576A..77S,2015A&A...576A..78K}.

\begin{acknowledgements}
NR acknowledges support by the DLR under project no. 50OR1002, a ``DAAD-Dok\-to\-randen\-sti\-pen\-di\-um'' as well as
the RTG 1351/2 "Extrasolar planets and their host stars". 
This research has made use of NASA's Astrophysics Data System as well as of the SIMBAD database and the VizieR catalogue access tool, operated at CDS, Strasbourg, France. We also want to thank an anonymous referee for exceptionally helpful reports and constructive criticism that helped to substantially improve the paper.
\end{acknowledgements}

\bibliographystyle{aa}
\bibliography{tellrem}

\end{document}